\documentclass[preprint,sort&compress,12pt]{elsarticle}

\usepackage[top=1in, bottom=1in, left=1in, right=1in]{geometry}

\usepackage{graphicx}
\usepackage{amsmath}
\usepackage{booktabs}
\usepackage{multirow}
\usepackage{natbib}
\usepackage{amssymb}
\usepackage{algorithm}
\usepackage{algpseudocode}
\usepackage{amsthm}

\usepackage{lineno}
\usepackage{subfig}
\usepackage{algorithm}
\usepackage{algpseudocode}
\usepackage{xcolor}
\usepackage[colorinlistoftodos]{todonotes}

\usepackage{xfrac}
\usepackage{acronym}
\acrodef{gan}[GAN]{\emph{GAN}}
\acrodef{bc}[BC]{\emph{BC}}
\acrodef{rbc}[RBC]{\emph{RBC}}

\biboptions{sort,compress} 

\usepackage{xcolor}

\newcommand*\patchAmsMathEnvironmentForLineno[1]{%
  \expandafter\let\csname old#1\expandafter\endcsname\csname #1\endcsname
  \expandafter\let\csname oldend#1\expandafter\endcsname\csname end#1\endcsname
  \renewenvironment{#1}%
     {\linenomath\csname old#1\endcsname}%
     {\csname oldend#1\endcsname\endlinenomath}}%
\newcommand*\patchBothAmsMathEnvironmentsForLineno[1]{%
  \patchAmsMathEnvironmentForLineno{#1}%
  \patchAmsMathEnvironmentForLineno{#1*}}%
\AtBeginDocument{%
\patchBothAmsMathEnvironmentsForLineno{equation}%
\patchBothAmsMathEnvironmentsForLineno{align}%
\patchBothAmsMathEnvironmentsForLineno{flalign}%
\patchBothAmsMathEnvironmentsForLineno{alignat}%
\patchBothAmsMathEnvironmentsForLineno{gather}%
\patchBothAmsMathEnvironmentsForLineno{multline}%
}

\usepackage{graphicx}
\usepackage{amssymb}
\usepackage{amsthm}
\usepackage{bbm}
\usepackage{bm}
\usepackage{lineno}
\usepackage{url}
\usepackage{listings}
\usepackage[colorlinks=true]{hyperref}

\usepackage{color,soul}
\usepackage{mathtools}

\definecolor{lightblue}{rgb}{.90,.95,1}
\definecolor{darkgreen}{rgb}{0,.5,0.5}

\definecolor{lightgreen}{rgb}{.90,1,0.90}

\usepackage{gensymb}
\usepackage{array}
\usepackage[commentmarkup=footnote]{changes}
\usepackage{multirow}
\usepackage{enumerate}
\newcolumntype{P}[1]{>{\centering\arraybackslash}m{#1}}

\newcolumntype{L}[1]{>{\raggedright\let\newline\\\arraybackslash\hspace{0pt}}m{#1}}
\newcolumntype{C}[1]{>{\centering\let\newline\\\arraybackslash\hspace{0pt}}m{#1}}
\newcolumntype{R}[1]{>{\raggedleft\let\newline\\\arraybackslash\hspace{0pt}}m{#1}}

\definechangesauthor[name={Heng Xiao}, color = purple]{hx}

\graphicspath{ {./} }

\usepackage{changes}
\definechangesauthor[name={Reviewer 1}, color = blue]{R1}
\definechangesauthor[name={Reviewer 2}, color = red]{R2}
\definechangesauthor[name={All reviewers}, color = brown]{All}
\definechangesauthor[name={Editor}, color = purple]{Editor}
\definechangesauthor[name={Author}, color = olive]{Author}

\begin{document}

\begin{frontmatter}


\title{First-principle-like reinforcement learning of nonlinear numerical schemes for conservation laws}

\author[stu]{Hao-Chen Wang}
\author[umbc]{Meilin Yu\corref{cor}}
\ead{mlyu@umbc.edu}
\author[stu]{Heng Xiao\corref{cor}}

\cortext[cor]{Corresponding authors}

\ead{heng.xiao@simtech.uni-stuttgart.de}

\address[stu]{Stuttgart Center for Simulation Science (SC SimTech), University of Stuttgart, Stuttgart, Germany}
\address[umbc]{University of Maryland Baltimore County (UMBC), Baltimore, MD, USA}

\begin{abstract}


Hyperbolic conservation laws governed by nonlinear partial differential equations (PDEs) pose grand challenges on numerical simulations as nonlinear numerical schemes with empirical model parameters, such as flux limiters, need to be designed to suppress spurious oscillations caused by numerical schemes near flow discontinuities and/or regions with under-resolved flow features. Unfortunately, there is no universal approach to design free-parameter-admitting nonlinear numerical schemes, and even if the model parameters can be determined following certain procedures, they still need to be tuned depending on the problems to be solved. 
In this study, we present a universal nonlinear numerical scheme design method enabled by multi-agent reinforcement learning (MARL). Different from contemporary supervised-learning-based and reinforcement-learning-based approaches, no reference data and special numerical treatments are used in the MARL-based method developed here; instead, a first-principle-like approach using fundamental computational fluid dynamics (CFD) principles, including total variation diminishing (TVD) and $k$-exact reconstruction, is used to design nonlinear numerical schemes.  
The third-order finite volume scheme is employed as the workhorse to test the performance of the MARL-based nonlinear numerical scheme design method. Numerical results demonstrate that the new MARL-based method is able to strike a balance between accuracy and numerical dissipation in nonlinear numerical scheme design, and outperforms the third-order MUSCL (Monotonic Upstream-centered Scheme for Conservation Laws) with the van Albada limiter for shock capturing. Furthermore, we demonstrate for the first time that a numerical scheme trained from one-dimensional (1D) Burger's equation simulations can be directly used for numerical simulations of both 1D and 2D (two-dimensional constructions using the tensor product operation) Euler equations. The working environment of the MARL-based numerical scheme design concepts can incorporate, in general, all types of numerical schemes as simulation machines. 

\end{abstract}


\end{frontmatter}

\setcounter{page}{2}

\nolinenumbers

\section{Introduction}

Hyperbolic conservation laws governed by nonlinear PDEs have widespread applications in many science and engineering fields, such as aero-hydrodynamics, astrophysics, plasma physics, advanced manufacture and transportation engineering~\cite{meister2012hyperbolic}. One feature of nonlinear hyperbolic conservation laws is that their solutions admit singularities (i.e., shock waves), which can be developed in finite time from smooth initial data. 
This poses grand challenges on numerical simulations as free-parameter-admitting nonlinear numerical schemes need to be developed to take both scheme stability and numerical resolution into consideration; see Godunov's pioneering work on  numerical methods for shock capturing~\cite{godunov1959finite}. As a result, many nonlinear numerical scheme construction methods have been developed during the last half century, such as high-resolution schemes with TVD slope limiters~\cite{van1979towards}, weighted essentially non-oscillatory (WENO) methods~\cite{harten1997uniformly,jiang1996efficient,cockburn1998essentially}, total variation bounded (TVB) discontinuous Galerkin methods~\cite{Cockburn_Shu_1989MC}, hierarchical multi-dimensional limiting process (MLP)~\cite{kim2005accurate,you2018high}, moving discontinuous Galerkin finite element method with interface condition enforcement (MDG-ICE)~\cite{corrigan2019moving,luo2021moving} and localized artificial viscosity and diffusivity methods~\cite{persson2006sub,yu2015localized,haga2019robust}, just to name a few. However, all nonlinear numerical schemes developed so far have to introduce empirical parameters, and even if these parameters can be determined following certain procedures, they still need to be tuned to achieve the scheme's best performance in different problems.   


\subsection{Supervised learning of shock capturing schemes}

To help ease the burden of parameter tuning, researchers began to use machine learning to design data-driven models. This type of model is different from the traditional model in the sense that most parameters which usually need to be tuned in the latter are incorporated into the supervised-learning based training process. For shock detection, Ray et al.~\cite{ray2018artificial,ray2019detecting} developed a data-driven troubled-cell indicator by training an artificial neural network (ANN) and tested the shock indicator on 1D grids and 2D unstructured grids. Beck et al.~\cite{beck2020neural} developed a data-driven shock indicator by using image-based edge detection methods on 2D grids. Numerical results show that both shock detectors perform better than the traditional ones and do not need problem-dependent parameter tuning.
However, one common feature shared by the aforementioned works is that they rely on the high-resolution reference data, which may not always be available, to train shock detectors due to the nature of supervised learning. 
Another issue is that special numerical treatment, such as certain auxiliary equations and their corresponding analytical solutions, needs be used to encode desired numerical features into the machine learning model. 

Apart from designing data-driven shock detection methods, researchers have also leveraged the tool of machine learning to design new flux limiters~\cite{nguyen2022machine} and learn discretizations for PDEs directly~\cite{bar2019learning}. Nguyen et al.~\cite{nguyen2022machine} designed a framework to derive an optimal flux limiter for the coarse-grained Burger's equation by learning from high-resolution data. The numerical results demonstrated that the trained flux limiter achieved better results than standard limiters in Burger's equation simulations.
We argue that the long-term numerical property (e.g., TVD) of the trained flux limiter may not be guaranteed, and the model generalizability is yet to be tested. Bar et al.~\cite{bar2019learning} designed a data-driven discretization method to learn the optimal approximations to PDEs on a coarse grid directly from the solutions on a finer grid. Numerical results demonstrated that their proposed method outperforms the standard numerical scheme on a coarse grid. However, the generalizability issues of the method was not discussed.
To address the intrinsic limitations associated with supervised learning, 
reinforcement learning provides an alternative way to directly learn strategies to build nonlinear numerical schemes without reference (labelled) data, thus increasing the chance to improve the model generalizability. 

\subsection{Reinforcement learning of shock capturing schemes}

Reinforcement learning aims at optimizing a control policy by maximizing the cumulative (discounted/delayed) reward and has a broad range of applications ranging from games~\cite{silver2017mastering,mnih2013playing}, robotics~\cite{zhao2020sim}, disease treatment~\cite{hu2023remedi}, and healthcare~\cite{yu2021reinforcement}. In computational fluid dynamics, reinforcement learning has been successfully applied to flow control~\cite{chen2023deep}, large eddy simulation (LES)~\cite{novati2021automating}, and wall-modelled LES~\cite{bae2022scientific}. Compared to supervised learning, labelled data are not needed. Instead, the agents in reinforcement learning repeatedly interact with the environment to generate interaction samples. 
In this way, the algorithm takes the response of the environment into account. 
This feature is of great importance because the long-term model properties can be guaranteed through the guide of a reward function which is carefully designed using expert or domain knowledge~\cite{sutton2018reinforcement}. Furthermore, if there exists a general strategy to solve a certain type of problems, e.g., a general numerical approach to solve special types of PDEs, reinforcement learning can then be used to identify this general strategy or policy.

Progress has been made to use reinforcement learning to design numerical schemes. Deep reinforcement learning was first used to learn WENO solvers for 1D scalar conservation laws via treating numerical PDE solvers as an Markov Decision Process (MDP) by Wang et al.~\cite{wang2019learning}. In their work, high-resolution solution data were included in the reward design. As a result, the agents were trained more or less in a data-driven or supervised fashion. No generalization of the trained numerical scheme to more complex equations, such as Euler equations, was demonstrated. The same limitation applies to the follow up work~\cite{way2022backpropagation,fu2022multi}.
Later, reinforcement learning was used to optimize the parameters of the fifth-order targeted ENO (TENO5)~\cite{fu2016family} for compressible flow simulations by Feng et al.~\cite{feng2023deep}. They designed a reward function that aims to optimize numerical dissipation and dispersion of the learned TENO5 scheme with reference to the fifth-order WENO and high-order central schemes. 

In an effort to improve generalizability, Beck and co-workers~\cite{schwarz2023reinforcement,keim2023reinforcement} designed slope limiters for 1D and 2D second-order finite volume schemes by using reinforcement learning without using high-resolution data for model training. The novelty of their works lies in that they included reference limiters in the reward design to bound the permissive slope in the time-stepping process. 
Specifically, they designed the reward function by following empirical rules of flux limiter design: (1) penalize near zero or negative density or pressure, (2) penalize oscillatory solutions, 
(3) reward solutions that are already the most non-oscillatory (corresponding to solution from the MinMod limiter), and (4) detect the upwind direction in nonshock solutions and utilize information derived from this.
They have demonstrated that the trained limiter can be generalized to solve the same governing equations with different initial conditions and grid sizes. However, a possible limitation of this design is that it relies on specific types of existing flux limiters, and thus, the trained nonlinear numerical schemes may inherit empirical elements therein.  As such, it is not clear how the learned scheme can be generalized to problems with different physics (e.g., from Burger's equation to Euler equations) and spatial dimensions (e.g., from one dimension to two and three dimensions), while standard numerical schemes in CFD typically have such generalization capability. 
The machine-learned numerical schemes should achieve a similar universality as in the standard CFD schemes across physics (i.e., governing equations), grid, and dimensionality.


\subsection{Objectives and contribution of the present work}

To achieve the goal of significantly enhancing the generalization capability of machine-learned numerical schemes, in this work we present a universal, first-principle-like nonlinear numerical scheme design method enabled by multi-agent reinforcement learning. This MARL-based numerical scheme design framework uses neither reference data nor empirical flux limiter features in the reward design. Instead, a first-principle-like approach is adopted, where fundamental CFD principles are used to guide the agents in seeking a balance between accuracy and numerical dissipation automatically.
Specifically, we design the reward function to promote stability by suppressing total variation increases with minimum possible intervention (i.e. using the limiter-free, $k$-exact finite volume reconstruction as the reference). Note that the use of a limiter-free scheme as a reference in reward design is fundamentally different from referring to a particular limiter~\cite{keim2023reinforcement}.
The trained nonlinear numerical scheme is largely free from \text{ad hoc} parameters and shows excellent generalizability over different physics, grid resolutions, and spatial dimensions. In doing so, a numerical scheme learned from the 1D Burger's equation is directly used to simulate both 1D and 2D Euler equations on grids with varying resolutions.

The rest of this paper is organized as follows. The MARL-based nonlinear numerical scheme is presented in Section~\ref{method}. Therein, the design of the equation environment, agents, and reward in MARL, and the associated training framework are introduced. In Section~\ref{sec:result}, numerical properties of the MARL-based nonlinear numerical schemes are examined, and test results of model generalizability across different types of governing equations, initial conditions, grid resolutions, and problem dimensions are presented and discussed. Finally, conclusions of the present work and discussions of potential improvement of the MARL-based nonlinear numerical scheme design method are provided in Section~\ref{conclusion}.

\section{Numerical Methodology}
\label{method}
\subsection{Multi-agent reinforcement learning}
Reinforcement learning is one of three basic machine learning paradigms, alongside supervised learning and unsupervised learning. The major difference between reinforcement learning and supervised learning is that the former algorithm does not need labelled input/output pairs. At each discrete time step $t$, with a given state $\mathbf{s}$, the agent selects actions $\mathbf{a}$ with respect to its policy $\pi:\mathcal{S} \rightarrow \mathcal{A}$, receiving a reward $r$ and the new state of the environment $\mathbf{s}'$. The return is defined as the discounted sum of rewards $R_t = \sum^{T}_{i=t} \gamma^{i-t}r(\mathbf{s}_i,\mathbf{a}_i)$, where $\gamma$ is a discount factor determining the priority of short-term rewards. The focus of reinforcement learning is to find an optimal policy $\pi_{\psi}$ that maximizes the expected return $J(\psi)$ through exploration and exploitation. 

There are different types of reinforcement learning algorithms. One simple but famous reinforcement learning algorithm is called Q-learning. In Q-learning, the algorithm has a function (which is usually called Q-function) that calculates the quality of a state-action combination: $Q:\mathcal{S} \times \mathcal{A}\rightarrow \mathbb{R}$. This quantity $Q$ in the expression is usually called Q-value. Q-learning seeks to find a policy that maximizes the expected return $J(\psi)$ through iteratively updating the Q-function. The limitation of such an algorithm is that only small-size problems with discrete action space can be handled.

However, with the traditional tabular solution replaced by a neural network solution, deep reinforcement learning is introduced. The addition of a function approximator like the neural network enables reinforcement learning to deal with very large input states and allow agents to make decisions from unstructured input data without manual engineering of the state space~\cite{sutton2018reinforcement}. Regardless of the additional effort needed for solving problems triggered by the use of a neural network, deep reinforcement learning is proved to be successful in different subjects such as games and robotics. Actually, the deep reinforcement learning can be viewed as the combination of a neural network expression and a reinforcement learning training scheme. One can view the interaction between the agents and the environment as a different way to generate the training data set used for the training of a neural network. 


In reinforcement learning, the number of agents can be more than one and this type of reinforcement learning is called multi-agent reinforcement learning. Compared to single-agent reinforcement learning, not only is the number of agents larger, but also agents can have different interests. In pure competition settings, the agents compete with each other. On the contrary, in pure cooperation settings, all the agents work together and get identical rewards. In other cases, the setting is usually the mix of both. For our problem, the setting is pure cooperation. Different agents pick different parameters based on different input states but work together to achieve a better long-term reward. To simplify the problem, only one policy (i.e., the actor neural network) is used. This simplification is consistent with the traditional design of a numerical scheme. The reward design will be discussed in Section~\ref{sec:reward}.


In this work, we use the multi-agent version of Twin-Delayed deep deterministic policy gradient~\cite{fujimoto2018addressing} (MATD3) and train our cooperative agents in a centralized training decentralized execution (CTDE) manner. MATD3 is able to address the problem of overestimation of Q-value by introducing three numerical treatments: clipped double-Q learning, delayed policy updates, and target policy smoothing. More details about the algorithm can be found in~\ref{sec:TD3}. 
In our implementation, all the agents share the parameters of a single neural network but take different actions based on different input states. The reinforcement learning processes are performed
using the open-source machine learning framework PyTorch~\cite{paszke2019pytorch} and the equation environments are implemented using the open-source Python Library Gym~\cite{brockman2016openai}.
\begin{figure}[!htbp]
  \centering
  \includegraphics[width=1.0\textwidth]{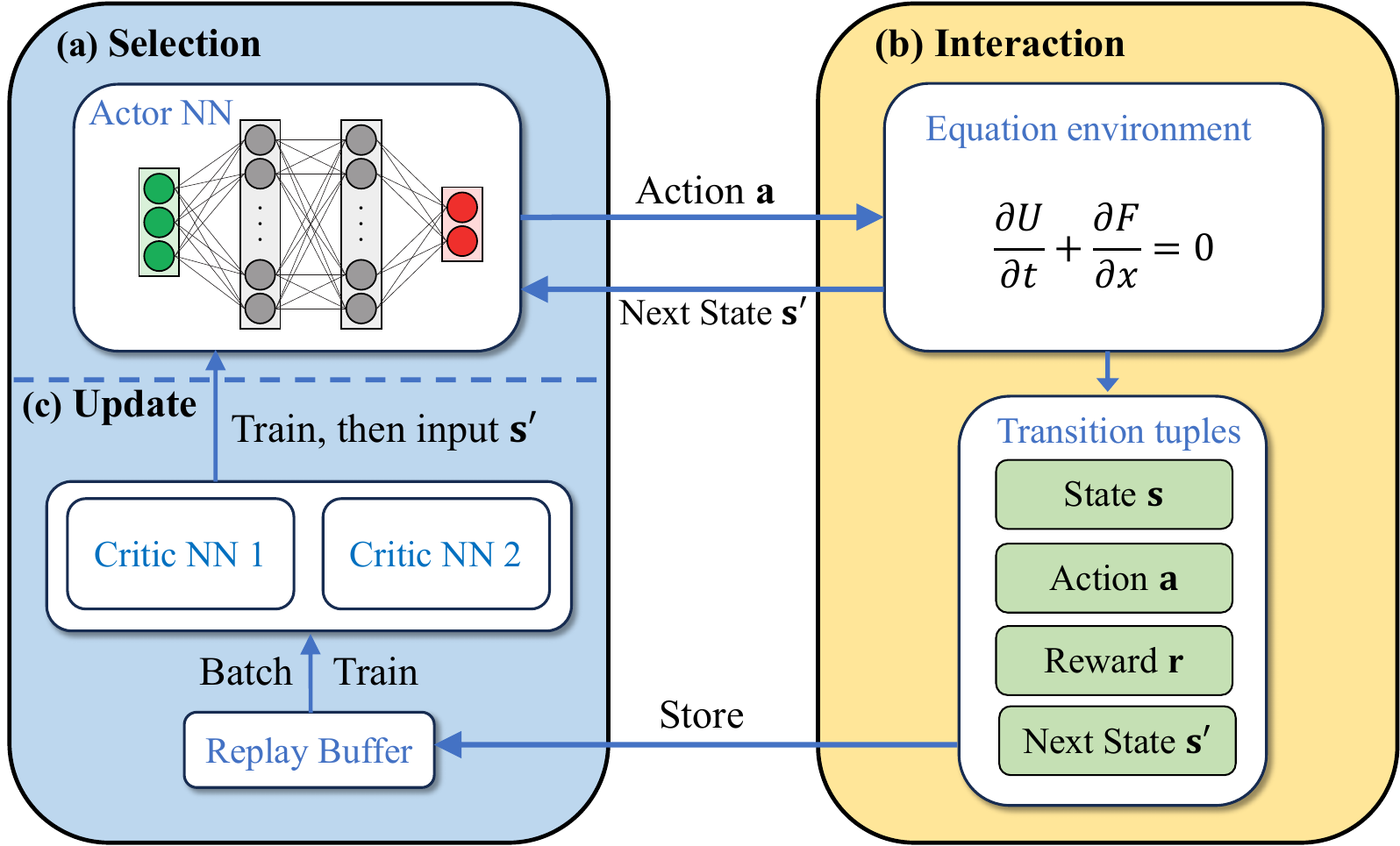}
    \caption{Workflow of the MARL-based training framework consisting of three steps: (a) input current states $\mathbf{s}$ to the actor neural network and output actions $\mathbf{a}$, (b) interact with the equation environment, generate numerous transition tuples, which are composed of current states $\mathbf{s}$, current actions $\mathbf{a}$, rewards $\mathbf{r}$, and next states $\mathbf{s'}$, and (c) store transition tuples in the replay buffer, batch samples from the replay buffer to update two critic neural networks first and then the actor neural network.}
  \label{fig:methodology}
\end{figure}

\subsection{MARL-based nonlinear numerical scheme}
In this section, we explain how to represent a nonlinear numerical scheme with the help of a neural network. Our nonlinear numerical scheme is built based on the MUSCL scheme. For a smooth problem (e.g., linear advection with sinusoidal wave initial condition), MUSCL can achieve the third-order accuracy by reconstructing the states ${u}_{i+\frac{1}{2}}^L$ and ${u}_{i+\frac{1}{2}}^R$ on the cell edge $i+1/2$ as follows:
\begin{subequations}   
\label{eq:muscl}
\begin{align}
{u}_{i+\frac{1}{2}}^L & = \bar{u}_{i}+\frac{\phi}{4}[(1 - \kappa)(\bar{u}_{i}-\bar{u}_{i-1})
+(1 + \kappa)(\bar{u}_{i+1}-\bar{u}_{i})],
\label{eqn1} \\
{u}_{i+\frac{1}{2}}^R & = \bar{u}_{i+1}-\frac{\phi}{4}[(1 - \kappa)(\bar{u}_{i+2}-\bar{u}_{i+1})
+(1 + \kappa)(\bar{u}_{i+1}-\bar{u}_{i})],
\label{eqn2}
\end{align}
\end{subequations}
where $\phi=1$ and $\kappa=1/3$, and $\bar{u}$ is the cell-averaged value.

However, for non-smooth problems, the best combination of $\phi$ and $\kappa$ remains unknown for different cells. Thus, we choose $\phi\in[0,2]$ and $\kappa \in[-1,1]$ to be our $i$th agent's actions $\mathbf{a}_{i} = \{a_{i,1},a_{i,2}\}$. The agents are distributed in each cell that needs to have these two parameters decided. This indicates that for a 1D problem the total number of agents is $N_T = N_x + 2$, where $N_x$ is the total number of cells and $2$ represents two ghost cells at each end. The input states $\mathbf{s}_{i}$ for the $i$th cell is a three-cell stencil which contains $\{{u}_{i-1}$, ${u}_{i}$, ${u}_{i+1}\}$ and is normalized by the variant min-max normalization, which reads
\begin{equation}
{\widehat{s}_{j}} = \Bigl\{
\begin{array}{lcl}
\frac{{s}_j - \min(\mathbf{s}_i)}{\max(\mathbf{s}_i) - \min(\mathbf{s}_i)} & \mbox{if} \; |\max(\mathbf{s}_i) - \min(\mathbf{s}_i)|> \epsilon_1\\
1 &  \mbox{otherwise}
\end{array}, \quad i=1,...,N, 
\end{equation}
where $\epsilon_1 = 10^{-8}$ is a small number.
We treat the nearly constant states that are caused by the numerical fluctuation as constant stencils and mark all the constant stencil since choosing different parameters has no effect on the numerical simulation results. 

The algorithm of the MARL-based nonlinear numerical scheme design framework is shown in Algorithm~\ref{alg:cap} and the hyperparameters are listed in Table~\ref{table:hyper}. The actor neural network (policy) ${\pi}_{\psi}$ as well as two critic networks $Q_{{\theta}_1}$, $Q_{{\theta}_2}$ all have 256 neurons in the hidden layers. There are also three corresponding target neural networks. All six networks have two hidden layers and are connected with the activation function ReLU~\cite{agarap2018deep}. 

We briefly explain Algorithm~\ref{alg:cap} here. As a first step, it initializes two critic networks and one actor network, as well as three target neural networks, with random parameters. A replay buffer $\mathcal{B}$ is also initialized for the storage of different transition tuples. Random action pairs are chosen for different agents to help agents explore the environment. Then a joint action is formed and used to forward the environment one time step using the second-order Runge–Kutta method. The environment then outputs the joint normalized next state and a joint reward. A transition tuple $(\mathbf{s}_i,\mathbf{a}_i,r_i,\mathbf{s}_i^{'})$ for each agent is stored in the replay buffer $\mathcal{B}$ afterwards. Then the algorithm batches $N_B$ transitions from $\mathcal{B}$ to train the critics networks and the actor neural network. The target neural networks are updated every $d$ steps. The actor neural network will output new action pairs $a_{i,1}^{t+1}, a_{i,2}^{t+1}$ based on ${\mathbf{s}_i}^{t+1}$. Then the iteration continues until reaching the end step of the environment. After that, the environment resets and begins to execute another training loop until meeting the final training step $n_{\max}$.
\begin{algorithm}[!htbp]
\caption{Reinforcement Learning Scheme (RLS) Framework}\label{alg:cap}
\begin{algorithmic}
\State Initialize critic networks $Q_{{\theta}_1}$, $Q_{{\theta}_2}$, 
and actor network ${\pi}_{\psi}$ with random parameters ${\theta_1}$, ${\theta_2}$, ${\psi}$.

\State Initialize target networks ${\theta}_1^{'} \leftarrow {\theta}_1$,
$\theta_2^{'} \leftarrow \theta_2$, 
${\psi^{'}} \leftarrow {\psi}$
and replay buffer $\mathcal{B}$

\While{$n=0$ $\leq$ $n_{\max}$}
    \State Initialize action pair $a_{i,1}, a_{i,2}$ for each agent
    \While {not $\mathbf{done}$} 
        \State Select action with exploration noise $\mathbf{a}_i \sim \mbox{clip} (\pi_{\psi}(\mathbf{s}_i)+\mathcal{N}(0,\sigma_{1}),\mathbf{a}\textsubscript{low},\mathbf{a}\textsubscript{high})$
        \State Compute $u_{i+1/2}^{L}$ and $u_{i+1/2}^{R}$ from Eqn.~\eqref{eq:muscl}
        \State Observe rewards $r_i$, new states $\mathbf{s}_i^{'}$ and termination signal $\mathbf{done}$ 
        for each agent
        \If {neither $\mathbf{s}_i$ nor $\mathbf{s}_i^{'}$ is constant}
            \State Store transition tuples $(\mathbf{s}_i,\mathbf{a}_i,r_i,\mathbf{s}_i^{'})$ in  $\mathcal{B}$
        \EndIf
        \If{$n$ mod $n\textsubscript{upfreq}$}
            \For{$l=1$ $\mathbf{to}$ $n\textsubscript{upfreq}$}
                \State Sample mini-batch of $N_B$ transitions $(\mathbf{s}_i,\mathbf{a}_i,r_i,\mathbf{s}_i^{'})$  from $\mathcal{B}$
                \State $\tilde{\mathbf{a}} \leftarrow \mbox{clip}  (\pi_{\psi'}(\mathbf{s}') + \mbox{clip}(\mathcal{N}(0,\sigma_{2}),-c,c),\mathbf{a}\textsubscript{low},\mathbf{a}\textsubscript{high})$
                \State $y \leftarrow r+\gamma \textrm{min}_{k=1,2}Q_{\theta_k^{'}}(\mathbf{s}',\tilde{\mathbf{a}})$
                \State Update critics $\theta_k \leftarrow \textrm{argmin}_{\theta_k}{N_B}^{-1} \sum(y-Q_{\theta_k}(\mathbf{s},\mathbf{a}))^2$
                \If {$n$ mod $d$}
                    \State Update $\psi$ by the deterministic policy gradient:
                    \State $\nabla_{\psi}J(\psi) = {N_B}^{-1} \sum \nabla_{a}Q_{\theta_1}(\mathbf{s},\mathbf{a})|_{\mathbf{a}={\pi}_{\psi}(\mathbf{s})}\nabla_{\psi}\pi_{\psi}(\mathbf{s})$
                    \State Update target networks:
                    \State $\theta_k^{'} \leftarrow \tau\theta_k + (1-\tau)\theta_k^{'}$
                    \State $\psi^{'} \leftarrow \tau\psi + (1-\tau)\psi^{'}$
                \EndIf
            \EndFor
        \EndIf
        
    \EndWhile
\EndWhile
\end{algorithmic}
\end{algorithm}

\begin{table}[h]
    \centering
    \begin{tabular}{lllll}
        \toprule
        Hyperparameters & Implementation \\
        \midrule
        Learning rate & $10^{-3}$  \\
        Optimizer & Adam~\cite{kingma2014adam} \\
        Target update rate ($\tau$) & $5 \times 10^{-3}$\\
        Batch size ($N_B$) & 1024\\
        Discount factor ($\gamma$) & 0.999 \\
        Normalized observation & True \\
        Exploration policy noise ($\sigma_{1}$) & $\mathcal{N}(0,0.2)$ \\
        Policy update frequency ($d$) & 2\\
        Policy noise ($\sigma_{2}$) & $\mathcal{N}(0,0.1)$\\
        Policy noise clip ($c$) & $2 \times 10^{-1}$\\
        Update frequency ($n\textsubscript{upfreq}$)& 200\\
        Max training step ($n_{\max}$)& $1 \times 10^{5}$\\
        \bottomrule
    \end{tabular}
\caption{Hyperparameters of the reinforcement learning algorithm.}
\label{table:hyper}
\end{table}

\subsection{Reward design}
\label{sec:reward}
In reinforcement learning, reward is of utmost importance. With a good presentation of the reward function, the agents can learn a desired control policy. In the context of the numerical scheme design, most previous research tends to include high-resolution data in the reward function to help the agents explore. This inclusion, however, makes the training be executed in a more supervised way because labelled data are used during the scheme training process. Taking the nature of reinforcement learning into consideration, we designed a reward function that does not require labeled data:
\begin{subequations}
\begin{equation}
r_i^t({\mathbf{s}}^{t+1}_i, {\mathbf{s}}^{t}_i,\mathbf{a}_i) =  \alpha \, r\textsubscript{D} + r\textsubscript{A},
\label{eq:reward}
\end{equation}
with
\begin{align}
r\textsubscript{D} & = \min(TV^{t} - TV^{t+1}, \; 0) ,
\label{eq:TV} \\
r\textsubscript{A} & = - ||\mathbf{a}_i - \mathbf{a}_\textsubscript{ref}||_{L_1} .
\label{eq:ra}
\end{align}
\end{subequations}
Herein, `$TV$' stands for the total variation defined as $TV = \sum_{i=1}^n|f(x_{i+1})-f(x_i)|$. Note that $f(x)$ is a non-dimensionalized variable; for example, when calculating the total variation of the fluid density $\rho$, $f(x)$ reads $\rho / \rho_{\infty}$, where $\rho_{\infty}$ is the density of the freestream serving as a non-dimensionalization reference.     $r\textsubscript{D}$ is a stability-promoting reward term, which suppresses the growth of total variation without actively promoting diffusion through a positive reward, i.e., $r\textsubscript{D}$ is always non-positive. $r\textsubscript{A}$ is an accuracy-promoting reward term, which measures in $L_1$-norm the deviation of the trained coefficients in a selected high-order accurate numerical scheme from their reference values. In this study, the third-order MUSCL scheme is selected as the high-order method template to design the accuracy-promoting reward term $r\textsubscript{A}$. As a result, $r\textsubscript{A}$ measures the deviation of the trained $\phi$ and $\kappa$ from their $k$-exact values in the MUSCL scheme, i.e., the reference values $\mathbf{a}_\textsubscript{ref} \equiv (\phi\textsubscript{ref}, \kappa\textsubscript{ref}) = (1, 1/3)$ as presented in Eqn.~\eqref{eq:muscl}, and thus, Eqn.~\eqref{eq:ra} can be written as $r\textsubscript{A} = -(|a_{i, 1} - 1|+|a_{i, 2} - 1/3|)$.

In Eqn.~\eqref{eq:reward}, $\alpha$ is a function of the Courant--Friedrichs--Lewy (CFL) number, which is used to blend the two parts of stability-promoting and accuracy-promoting rewards. In this study, $\alpha$ is defined as $\alpha = \alpha_0/\text{CFL}$, where $\alpha_0$ is a constant. A key consideration to build $\alpha$ is to ensure that the reward can strike a good balance between adding numerical dissipation to ensure stability and maintaining high-order numerical scheme features to promote accuracy. Note that the maximum magnitude of $r\textsubscript{A}$ in our study is $7/3$ when $\phi$ is $0$ or $2$, and $\kappa$ is $-1$. 
To match the order of magnitude of $r\textsubscript{A}$, the term $\alpha \, r\textsubscript{D} = \alpha_0 \, r\textsubscript{D}/\text{CFL}$ should have an order of magnitude of $\mathcal{O} (1)$. Through numerical experiments, we find that the machine-learned numerical schemes are not sensitive to the variation of $\alpha_0$ when it varies between 1 and 100. Therefore, $\alpha_0$ is set as 50 in this study.
More details regarding the choice of the parameter $\alpha_0$ are discussed in~\ref{sec:parameter}. 

\begin{figure}[!htbp]
  \centering
  \subfloat[Results from stability-promoting scheme]
  {\includegraphics[width=0.48\textwidth]{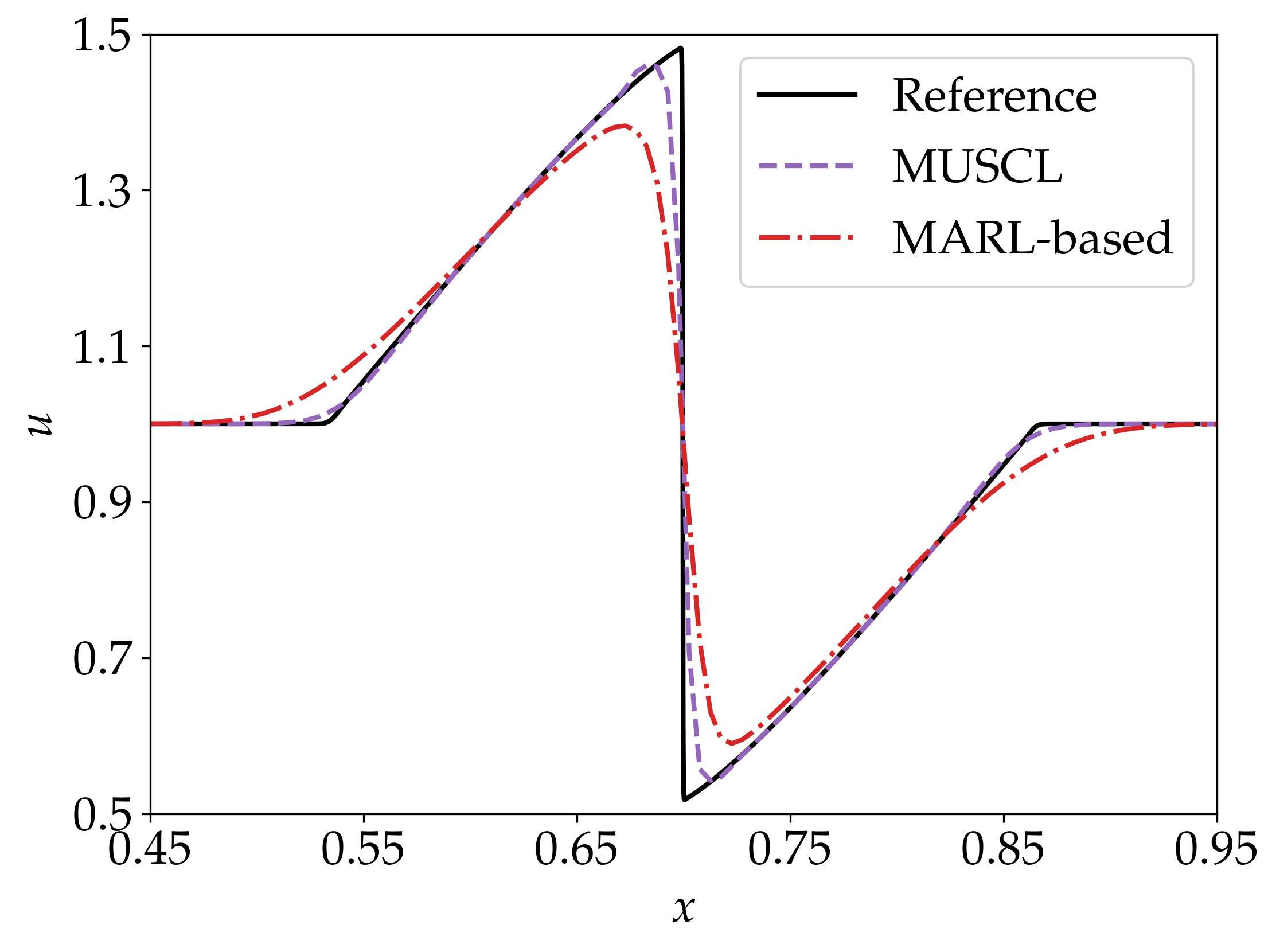}}\hspace{0.2em}
\hspace{0.2em}
  \subfloat[Results from accuracy-promoting scheme]
  {\includegraphics[width=0.48\textwidth]{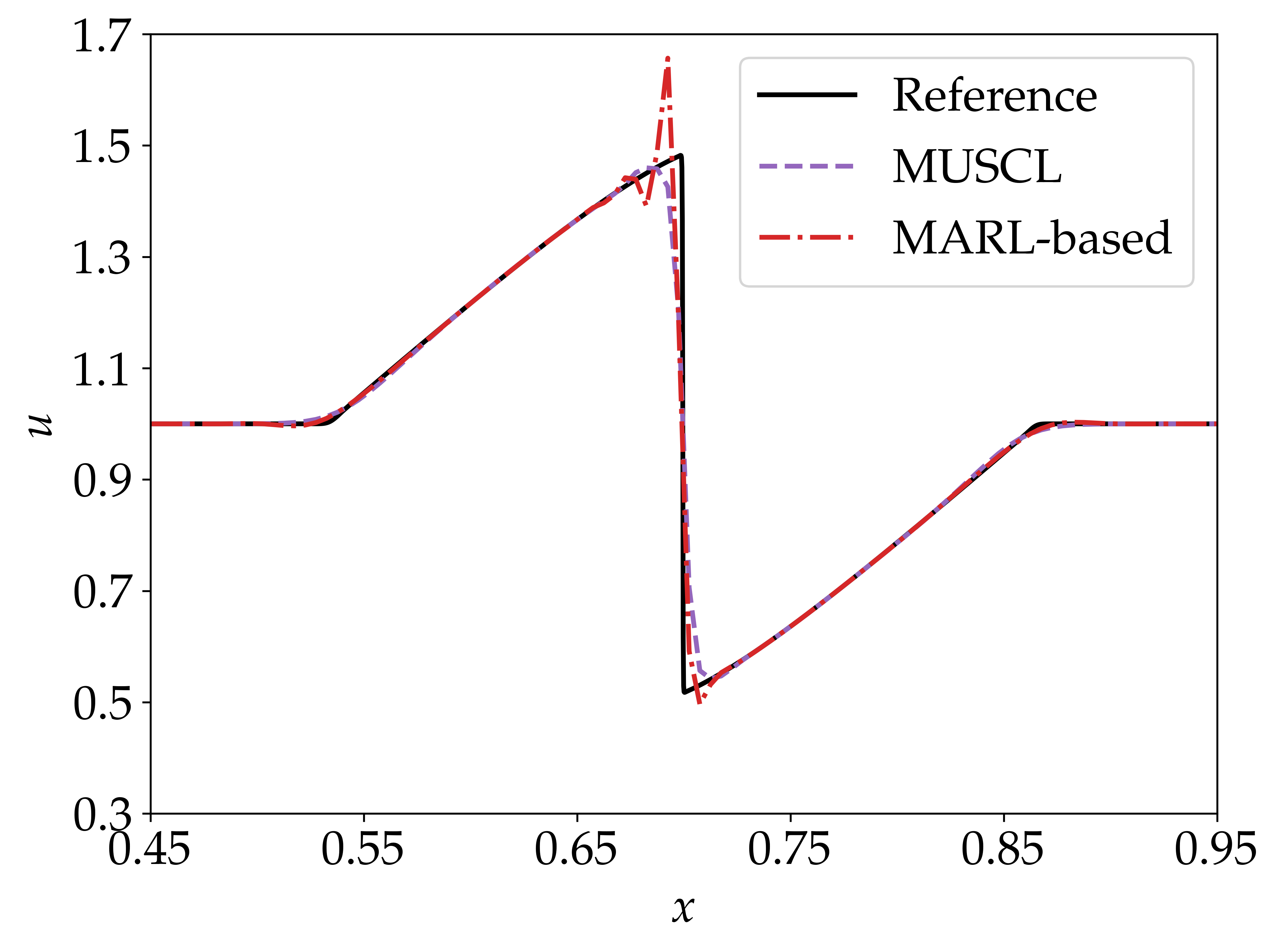}}
    \caption{Numerical results of the 1D inviscid Burger's equation with sinusoidal wave initial condition generated with (a) only the first part of the reward $r\textsubscript{D}$ being active, and (b) only the second part of the reward $r\textsubscript{A}$ being active.}
  \label{fig:burger0}
\end{figure}

The stability-promoting term $r\textsubscript{D}$ and the accuracy-promoting term $r\textsubscript{A}$ play different roles in the reward function.
Equation~\eqref{eq:TV} is designed to control the total variation of the numerical scheme to smear non-physical oscillations. If the total variation $TV^{t+1}$ of the next state $\mathbf{s}^{t+1}$ is larger than the $TV^{t}$ of the current state $\mathbf{s}^{t}$, a negative reward is distributed to agents who contribute to the increase of total variation. In Fig.~\ref{fig:burger0}a, we demonstrate the simulation results of the 1D Burger's equation with only the first part of the reward $r\textsubscript{D}$ being activated. The result is very dissipative compared to the reference solution because the algorithm simply chooses parameters that will make the total variation diminish.

In Eqn.~\eqref{eq:ra}, the algorithm compares the chosen actions at every time step to the $k$-exact (i.e., $2$-exact, third-order in our study) reconstruction parameters for smooth problems as presented in Eqn.~\eqref{eq:muscl} to dynamically adjust the actions along the simulation process to make the scheme less dissipative. In Fig.~\ref{fig:burger0}b, we demonstrate the simulation results of the 1D Burger's equation with only the second part of the reward $r\textsubscript{A}$ being activated. There exist expected oscillations near the shock because the trained numerical scheme can be considered as a MUSCL scheme with low artificial numerical dissipation. Note that the control policy in reinforcement learning can add default numerical dissipation to the trained nonlinear numerical scheme; otherwise, numerical simulations will blow up due to Gibbs oscillations and the reward will be penalized correspondingly.

Together with both parts of the reward activated, the training algorithm is searching to strike a balance between numerical dissipation and accuracy. Based on our reward design, there is no need of including any high-resolution solution data as references or empirical flux limiter features. Instead, we design the reward only based on a first-principle-like approach using fundamental CFD principles: control the total variation of numerical solutions to achieve better numerical stability and control the numerical dissipation by using $k$-exact reconstruction parameters for smooth problems as a guidance to achieve better accuracy. The agents, with the help of reinforcement learning, are able to find a balance between numerical dissipation and accuracy through continuous exploration and exploitation of interaction samples in the equation environments.

\section{Results and Discussions}
\label{sec:result}
In this section, the capability of MARL-based nonlinear numerical scheme is demonstrated with several sets of shock-capturing problems governed by the 1D inviscid Burger's equation, 1D Euler equations, and 2D Euler equations. They are all representative cases of hyperbolic conservation laws. The schemes presented here are trained only from the 1D inviscid Burger's equation with a sinusoidal wave initial condition and with a shock developed at the final simulation stage. The generalizability of the trained 1D nonlinear numerical scheme is tested with a widespread of cases of different physics, grid
resolutions, and spatial dimensions. 

\subsection{1D inviscid Burger's equation}
\label{sec:burger}
The 1D inviscid Burger's equation is a fundamental scalar hyperbolic conservation law that can develop discontinuities in finite time. It reads
\begin{equation}
\frac{\partial u}{\partial t} + \frac{\partial (u^2/2)}{\partial x} = 0,
\end{equation}
where $u$ is the working variable of Burger's equation.
The sinusoidal wave initial condition is given as follows,
\begin{equation}
u = \Bigl\{
\begin{array}{lcl}  1+\sin(6\pi(x-1/3))/2 &\mbox{if} & 1/3\le x\le 2/3\\
1 &\mbox{if}  & x<1/3 \; \mbox{or} \; 2/3<x
\end{array}.
\end{equation}
The simulation parameter setting for this problem is defined as follows: the number of cells within the computational domain is $N_x=200$, the simulation end time is $T_{\textrm{end}} = 0.2$, and the CFL number is fixed at $0.2$.

We use the proposed MARL-based design method Algorithm~\ref{alg:cap} to train nonlinear numerical schemes with the 1D inviscid Burger's equation, and choose two typical schemes with different numerical properties to demonstrate the performance. The first scheme prefers numerical dissipation and is called RLS-D, in which RLS is the acronym of reinforcement learning scheme. The second scheme prefers numerical accuracy and we refer to this scheme as RLS-A. 

\begin{figure}[!htbp]
  \centering
  \subfloat[Numerical results]{\includegraphics[width=0.48\textwidth]{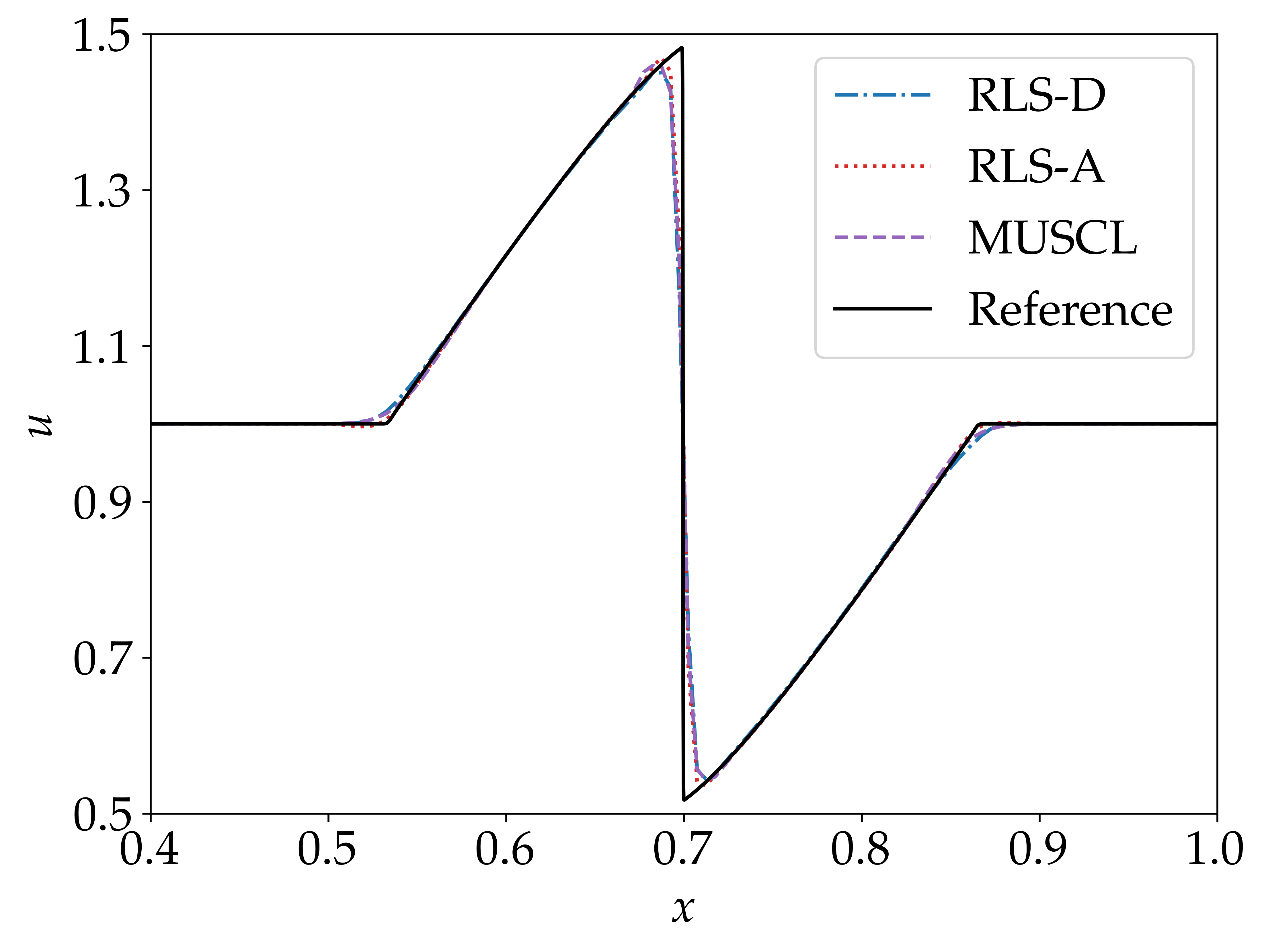}}\hspace{0.2em}
\hspace{0.2em}
  \subfloat[Total variation change]{\includegraphics[width=0.48\textwidth]{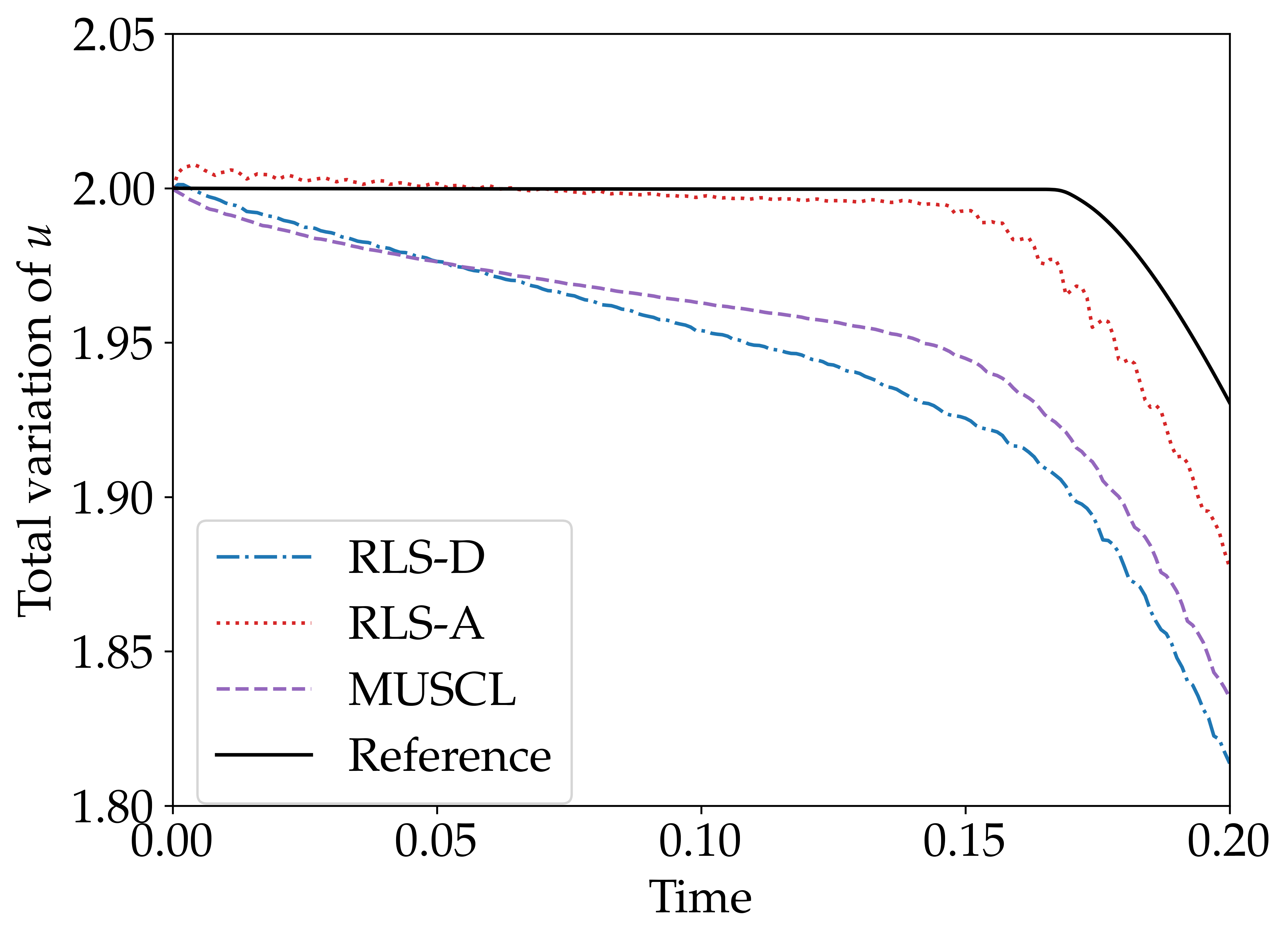}}
    \caption{(a) Comparison of numerical simulation results of the 1D inviscid Burger's equation with sinusoidal wave initial condition generated from RLS-D, RLS-A, the third-order MUSCL scheme with van Albada limiter and reference solution. (b) Total variation changes of MARL-based nonlinear numerical schemes.}
  \label{fig:burger1.1}
\end{figure}

The results in Fig.~\ref{fig:burger1.1}a show that numerical results generated from both schemes have good agreement with reference solution data as the shock is well captured even though in reward design we did not include any high-resolution data. We also include the numerical result calculated from the third-order MUSCL scheme using the van Albada limiter with the same grid size for comparison. Note that the reference solution is calculated from the same MUSCL scheme but on a very fine mesh in which the number of cells is $N_x=6400$. We find from Fig.~\ref{fig:burger1.1}a that RLS-A slightly outperforms RLS-D and the third-order MUSCL. 
In Fig.~\ref{fig:burger1.1}b, we observe that total variation is under good control throughout the entire simulation. Both RLS-A and RLS-D are able to retrieve the TVD numerical property successfully. For RLS-A, before the shock formation, its total variation decreases much slower than RLS-D, which indicates that the scheme is able to choose actions that are less dissipative during the smooth wave propagation. After the shock is formed, the scheme switches its actions to a dissipative mode.

\subsection{Generalization across different types of governing equations}
\label{sec:euler}
Euler equations are a set of equations that describe the flow of inviscid, compressible fluids. In 1D form, Euler equations for an ideal gas consist of three equations that govern the conservation of mass, momentum and energy. They are written in the vector format as:

\begin{equation}
\frac{\partial U}{\partial t} + \frac{\partial F}{\partial x} = 0,
\label{eq:Euler_1D}
\end{equation}
with
\begin{eqnarray*}
U  = 
\left (
\begin{array}{c}
      \rho   \\
      \rho u  \\
      E
\end{array} \right ), \quad \text{and} \quad
F =
\left (
\begin{array}{c}
      \rho u  \\
      \rho u^2 + p  \\
      u (E+p)
\end{array} \right ).
\end{eqnarray*}

Herein, $\rho$ is the density, $u$ is the velocity, $p$ is the pressure, and $E$ is the total energy. The ideal gas law $E=p/(\gamma_g-1) + \rho u^2/2$, where $\gamma_g = 1.4$ is the constant specific heat ratio, is used to close the system. 
The Sod shock tube problem is used to conduct tests in this section, and its initial condition is given as follows,
\begin{equation}
    (\rho,\, u,\, p) = \Bigl\{
    \begin{array}{lcl}(1,\,0,\,1) & x<0.5\\
    (0.125,\,0,\,0.1) & x\ge0.5
    \end{array}.
\end{equation}

\begin{figure}[!htbp]
  \centering
  \subfloat[Numerical results]{\includegraphics[width=0.48\textwidth]{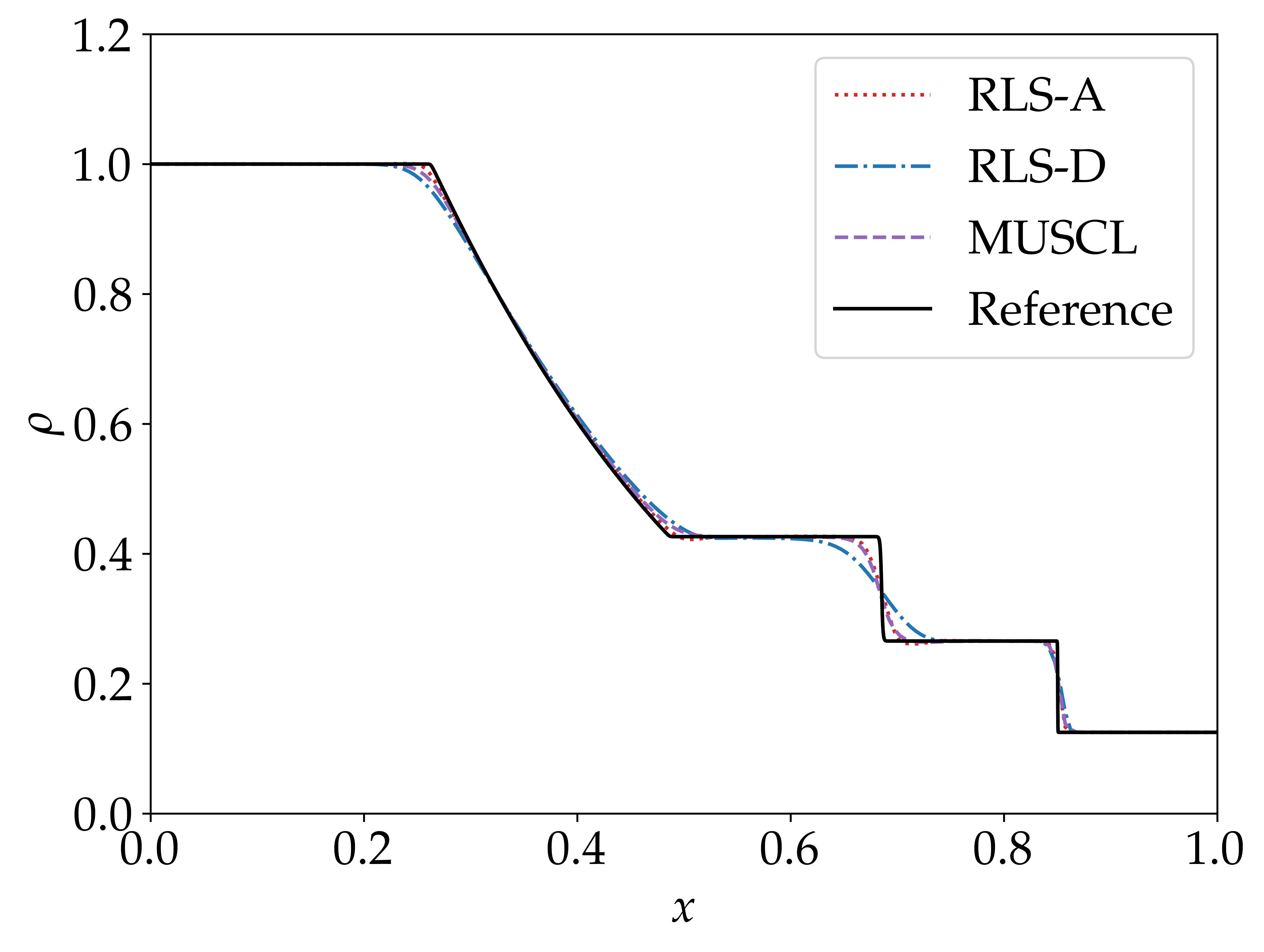}}\hspace{0.2em}
\hspace{0.2em}
  \subfloat[Total variation change]{\includegraphics[width=0.48\textwidth]{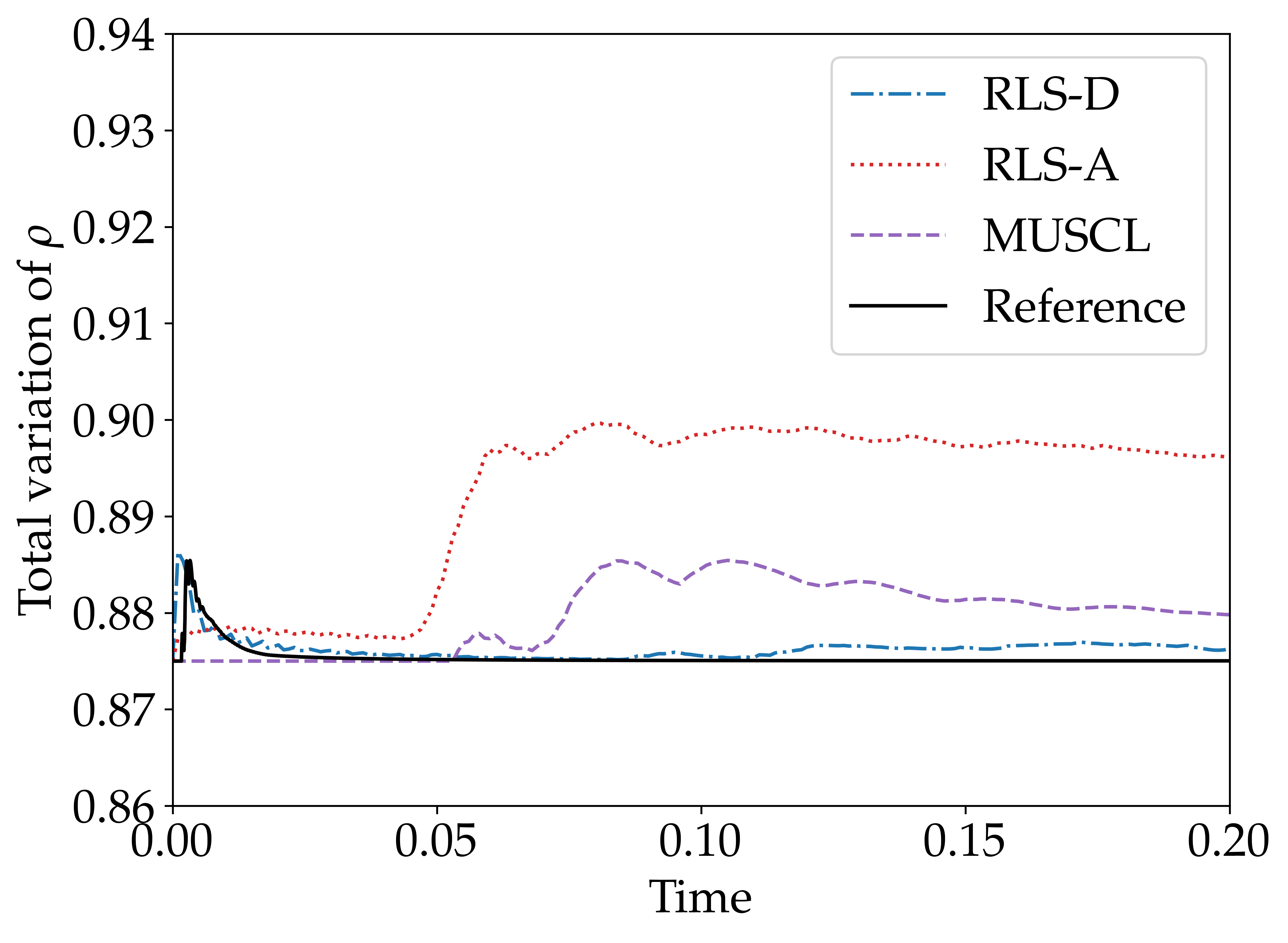}}
    \caption{(a) Comparison of numerical simulation results of density $\rho$ of the 1D Euler equations with Sod shock tube initial condition generated from RLS-D, RLS-A, the third-order MUSCL scheme with van Albada limiter and reference solution. (b) Total variation changes of MARL-based nonlinear numerical schemes}
  \label{fig:euler1}
\end{figure}
The simulation parameter setting for this problem is defined to be the same as that of the Burger's equation: $N_x = 200$, $T_{\textrm{end}} = 0.2$, and the CFL number is $0.2$. The reference solution is calculated using the same MUSCL scheme on a finer grid, where the number of cells is $N_x=6400$.

To demonstrate the generalizability of MARL-based nonlinear numerical schemes to different physics, we use the same schemes trained from the 1D inviscid Burger's equation to directly solve the 1D Euler equations. 
In this study, we use the local Lax-Friedrichs (i.e., Rusanov) method to solve the Riemann problem in the 1D Euler equations. We observe from Fig.~\ref{fig:euler1}a that without any prior knowledge of Euler equations or the assistance of high-resolution solution data, numerical results of density $\rho$ from machine-learned schemes have good agreement with those from the MUSCL scheme on the same grid and reference solution. RLS-D underperforms the third-order MUSCL scheme in the contact discontinuity region, i.e., showing more dissipative results there, due to that the scheme prefers adding more numerical dissipation to stabilize the simulation. In contrast, RLS-A is capable of capturing the contact discontinuity sharper than RLS-D and the third-order MUSCL theme. 
In Fig.~\ref{fig:euler1}b, we observe that total variations of the density from RLS-A and RLS-D are under good control. Note that the total variation of the Sod shock tube problem should remain constant during the development of the rarefaction wave, contact discontinuity, and shock wave. Any total variation increase indicates the creation of local extremes. From an enlarged view near the end of the rarefaction wave (i.e., $x \approx 0.5$), and the contact discontinuity (i.e., $x \approx 0.7$) in Fig.~\ref{fig:euler1}a, we observe apparent, although very small, troughs from the density predicted by RLS-A. This is due to the low numerical dissipation nature of RLS-A, and explains why the total variation value of RLS-A is larger than that of the reference value. 

In summary, the good generalizability across Burger's equation and Euler equations demonstrated in this section is consistent with standard CFD schemes, and has not been demonstrated in previous reinforcement learning based works.

\subsection{Generalization across different initial conditions}
\label{sec:euler_shu}
\begin{figure}[!htbp]
  \centering
  \subfloat[Numerical results]{\includegraphics[width=0.48\textwidth]{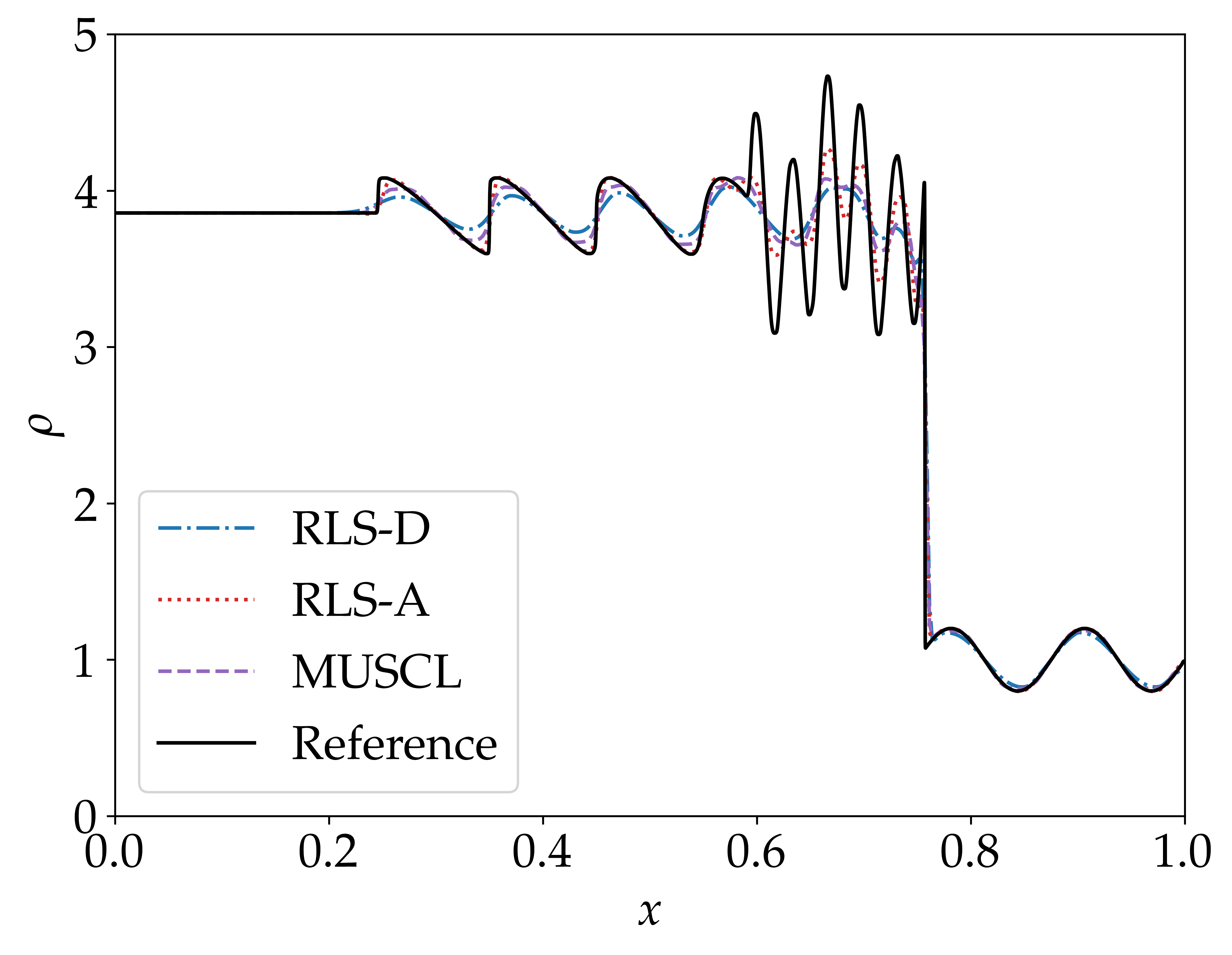}}\hspace{0.2em}
\hspace{0.2em}
  \subfloat[Total variation change]{\includegraphics[width=0.48\textwidth]{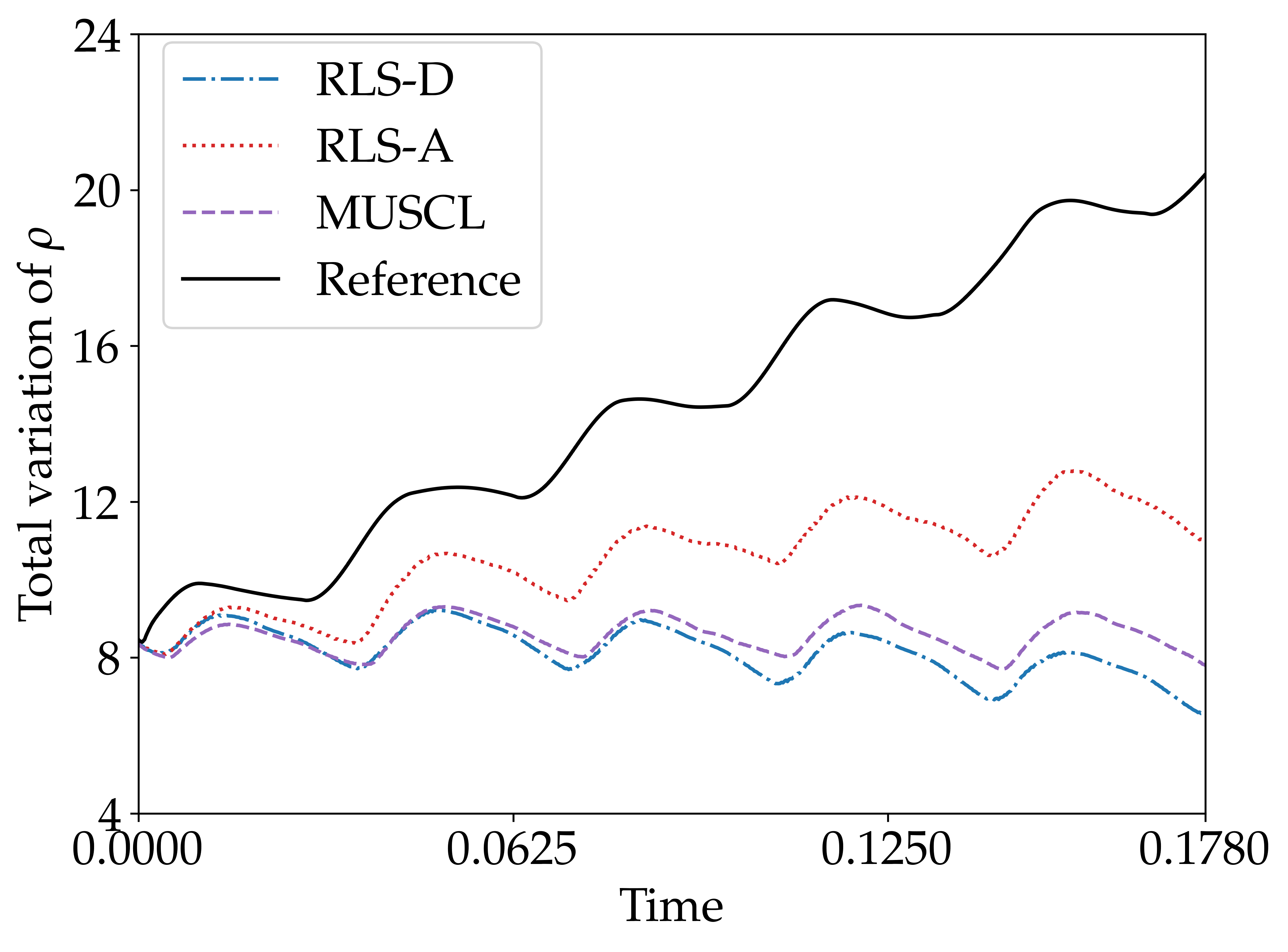}}
    \caption{(a) Comparison of numerical simulation results of density $\rho$ of the 1D Euler equations with Shu-Osher shock tube initial condition generated from RLS-D, RLS-A, the third-order MUSCL scheme with van Albada limiter and reference solution. (b) Total variation changes of each solutions. 
    }
  \label{fig:euler2.1}
\end{figure}

In this section, the same scheme trained from the 1D Burger's equation is tested with the 1D Shu-Osher problem where a shock wave interacts with an entropy wave. Its initial conditions are given as:
\begin{equation}
    (\rho,\, u,\, p) = \Bigl\{
    \begin{array}{lcl}(3.857143,\,2.629369,\,10.3333) & x<1/8\\
    (1+0.2\sin(16\pi x),\,0,\,1) & x\ge1/8
    \end{array}.
\end{equation}
The computational domain is $[0,1]$, and a grid with $N_x=400$ cells is used to conduct simulations. The CFL number is set as $0.1$ and the end time is set to $T_{\textrm{end}} = 0.178$. The reference solution is calculated using the same MUSCL scheme on a finer grid, where the number of cells is $N_x=6400$. Similar to the Sod shock tube test presented in Section~\ref{sec:euler}, we observe in Fig.~\ref{fig:euler2.1}a that RLS-A outperforms RLS-D and MUSCL, and RLS-D is the most dissipative scheme among the three. For all schemes, the amplitudes of the high-frequency waves are under-predicted compared to the reference solution. Total variations of the density from all tests are presented in Fig.~\ref{fig:euler2.1}b. It is interesting to observe that RLS-D is TVD in a time-averaged sense, and this corresponds well with the dissipative nature of RLS-D. Note that in the shock-entropy wave interaction, new extremes can be created. As a result, the total variation of the conserved variables can increase (see the reference solution). 
The time-averaged total variation of RLS-A captures this feature although its increasing rate is smaller than that of the reference solution.

\subsection{Generalization across varying grid resolutions}
\label{sec:grstudy}
In this section, we present the numerical results of grid refinement studies for the Sod problem and Shu-Osher problem to test the generalizability of the MARL-based schemes across varying grid resolutions. Specifically, the grid is refined from a coarse one with 100 cells to a fine one with 800 cells, and both RLS-A and RLS-D are tested. 
\begin{figure}[!htbp]
  \centering
  \subfloat[Density RLS-D]
  {\includegraphics[width=0.48\textwidth]{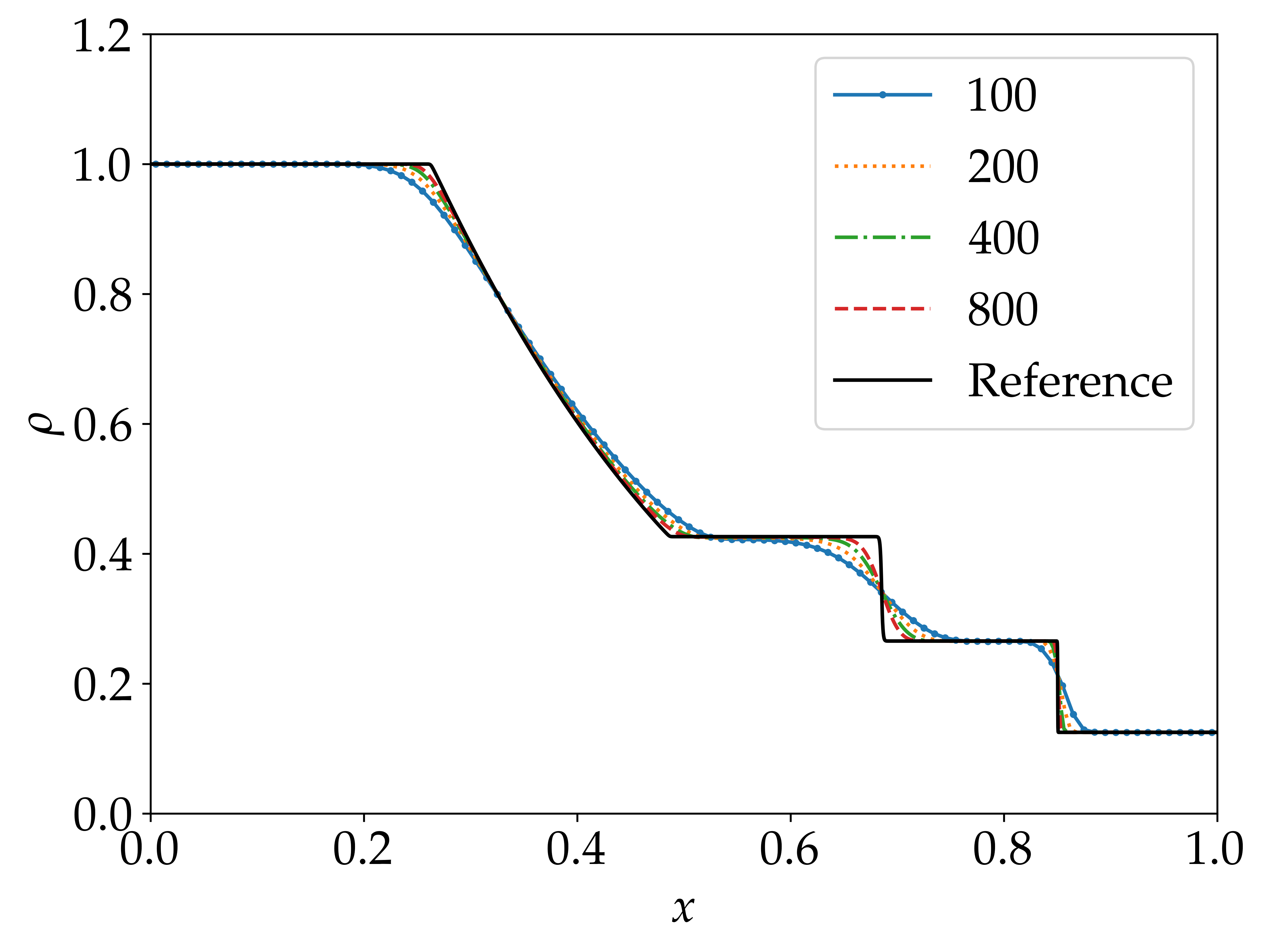}}\hspace{0.2em}
\hspace{0.2em}
  \subfloat[Density RLS-A]
  {\includegraphics[width=0.48\textwidth]{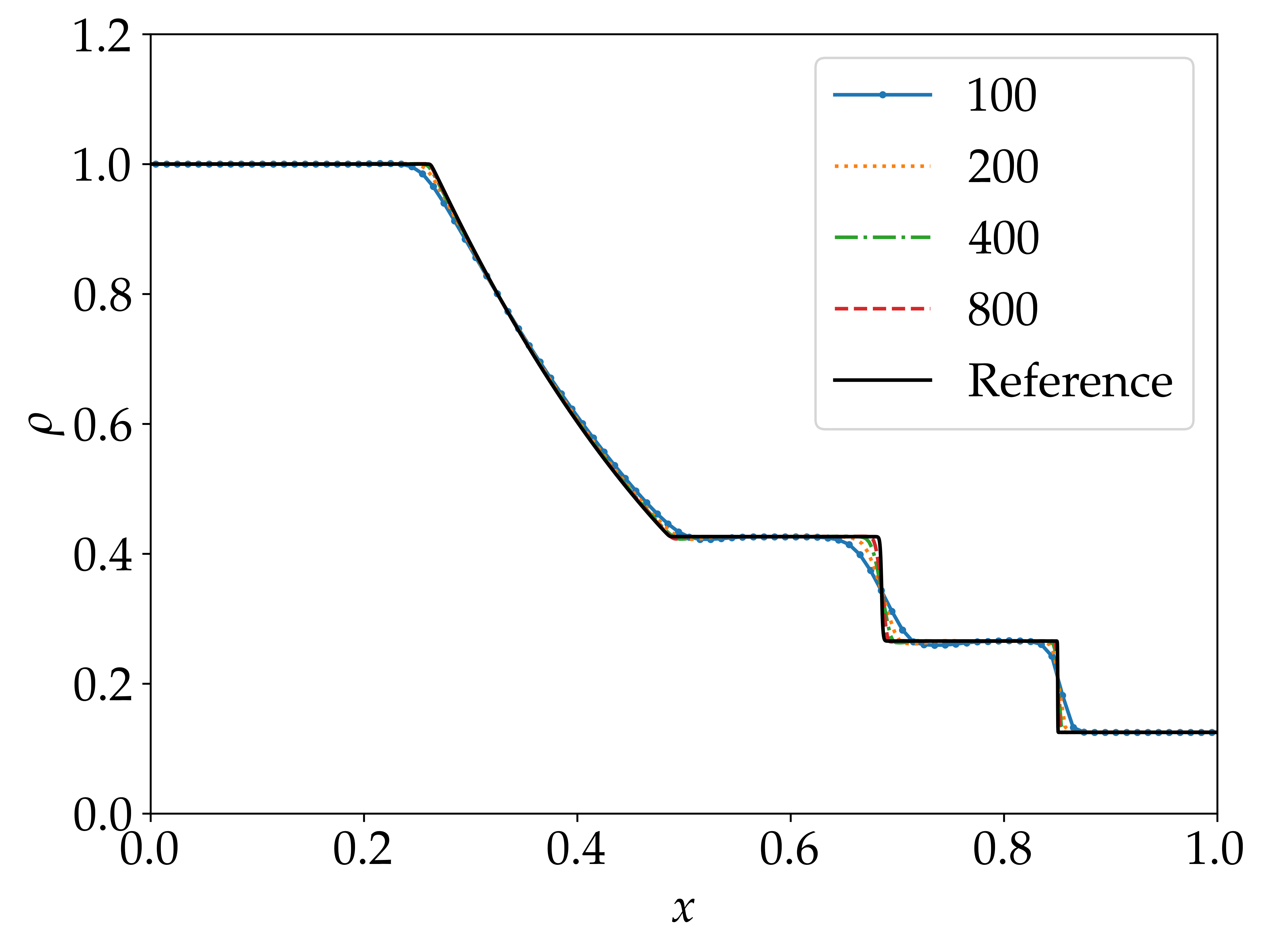}}\hspace{0.2em}
  \hspace{0.2em}
  \subfloat[TV RLS-D]{\includegraphics[width=0.48\textwidth]{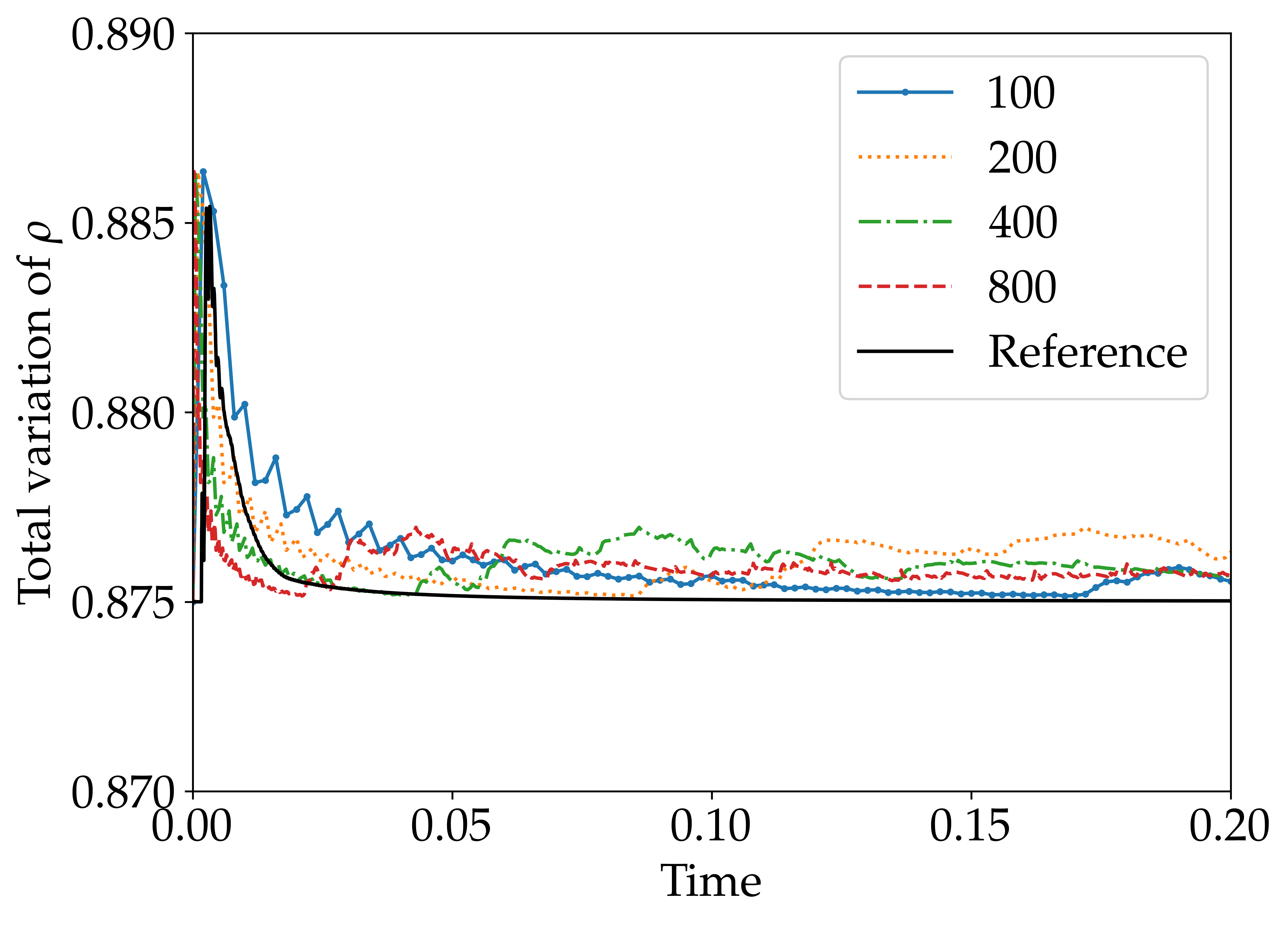}}\hspace{0.2em}
\hspace{0.2em}
  \subfloat[TV RLS-A]{\includegraphics[width=0.48\textwidth]{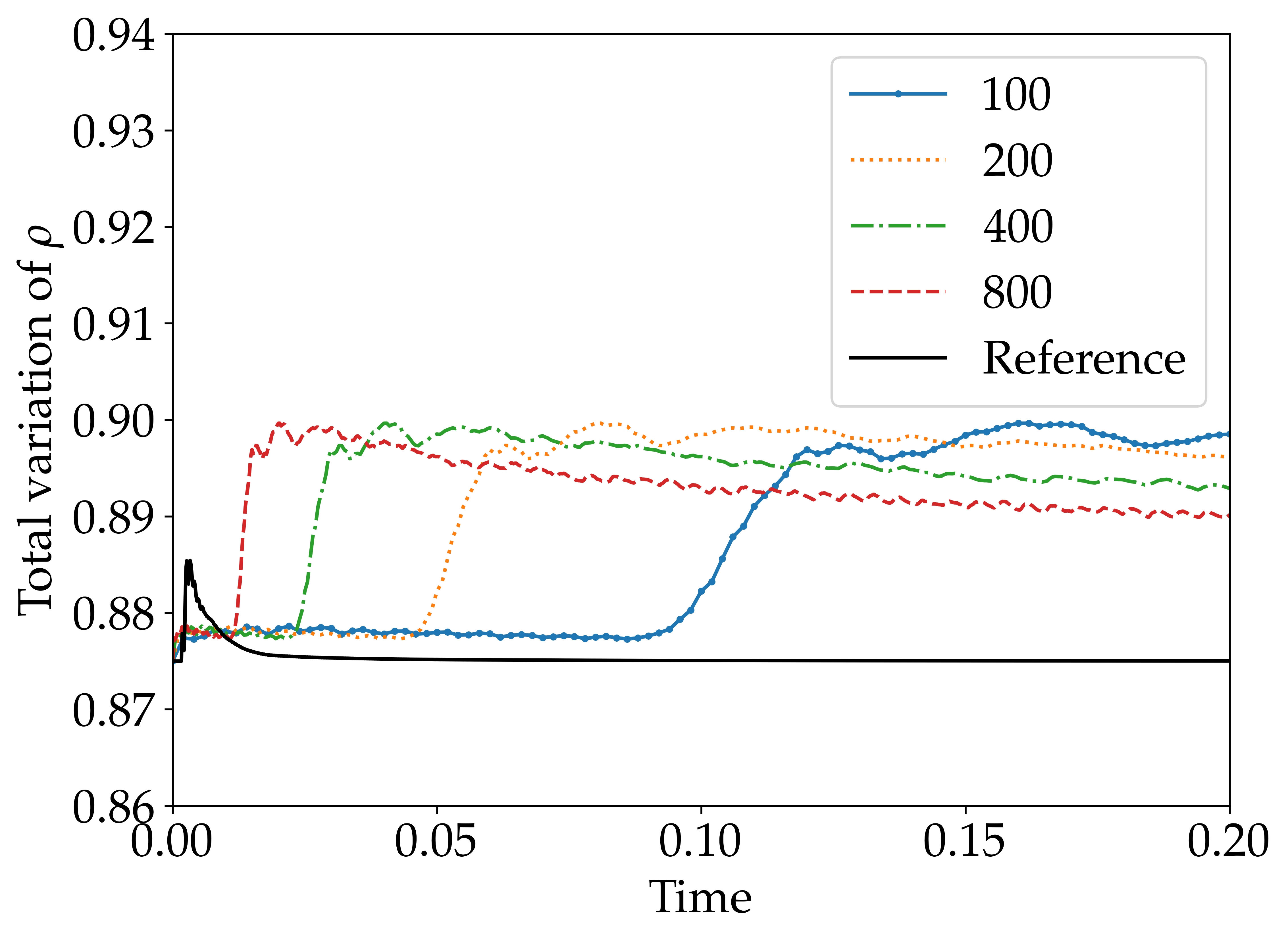}}
    \caption{Grid refinement study results of the 1D Euler equations with the Sod shock tube initial condition. (a) Density fields for RLS-D, (b) density fields for RLS-A, (c) total variation evolution histories for RLS-D, and (d) total variation evolution histories for RLS-A are presented here for comparison.}
  \label{fig:euler3.1}
\end{figure}

We observe in Fig.~\ref{fig:euler3.1} that when the grid is refined, density fields of the Sod shock tube problem from both schemes become more resolved and accurate. As expected, RLS-A performs better than RLS-D as it can capture the rarefaction wave, contact discontinuity, and shock wave with sharper resolutions. We observe from Fig.~\ref{fig:euler3.1}a that RLS-D can sharply capture the shock wave when the grid is refined to 400 cells. However, even with 800 cells, RLS-D cannot sharply capture the contact discontinuity. Instead, RLS-A with 800 cells can almost achieve the resolution of the reference solution with 6400 cells. We find from Fig.~\ref{fig:euler3.1}c that the total variation evolution histories of RLS-D with different grid resolutions do not show large differences when the wave propagation reaches a relatively stable stage, i.e., rarefaction, contact discontinuity, and shock have been sufficiently separated from each other.  From Fig.~\ref{fig:euler3.1}d, we find that when the grid is refined, the total variation evolution history of RLS-A is converging towards the reference one. This trend corresponds well to the decreasing local extreme values near the end of the rarefaction wave and the contact discontinuity as observed in Fig.~\ref{fig:euler3.1}b.

\begin{figure}[!htbp]
  \centering
  \subfloat[RLS-D]{\includegraphics[width=0.48\textwidth]{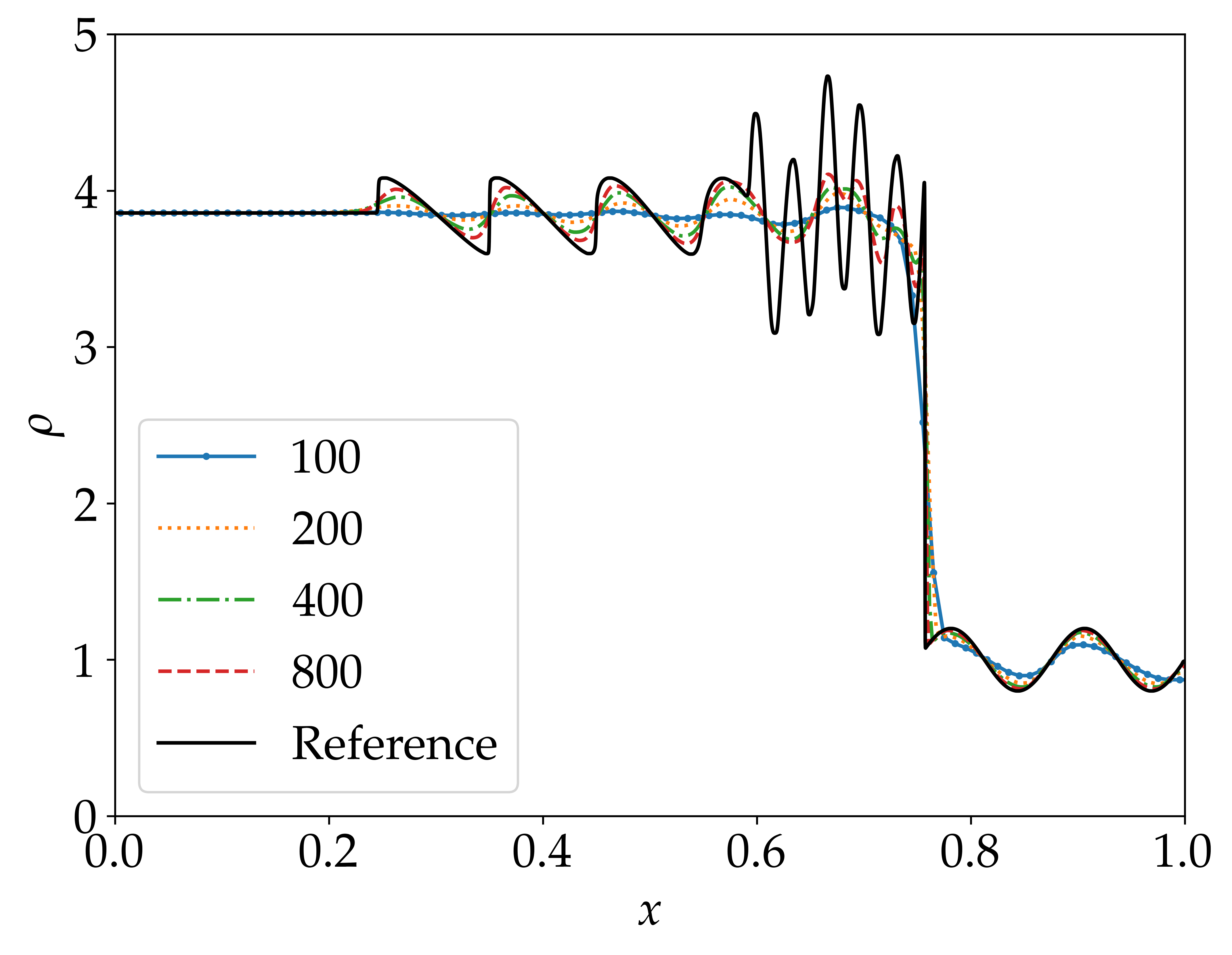}}\hspace{0.2em}
\hspace{0.2em}
  \subfloat[RLS-A]{\includegraphics[width=0.48\textwidth]{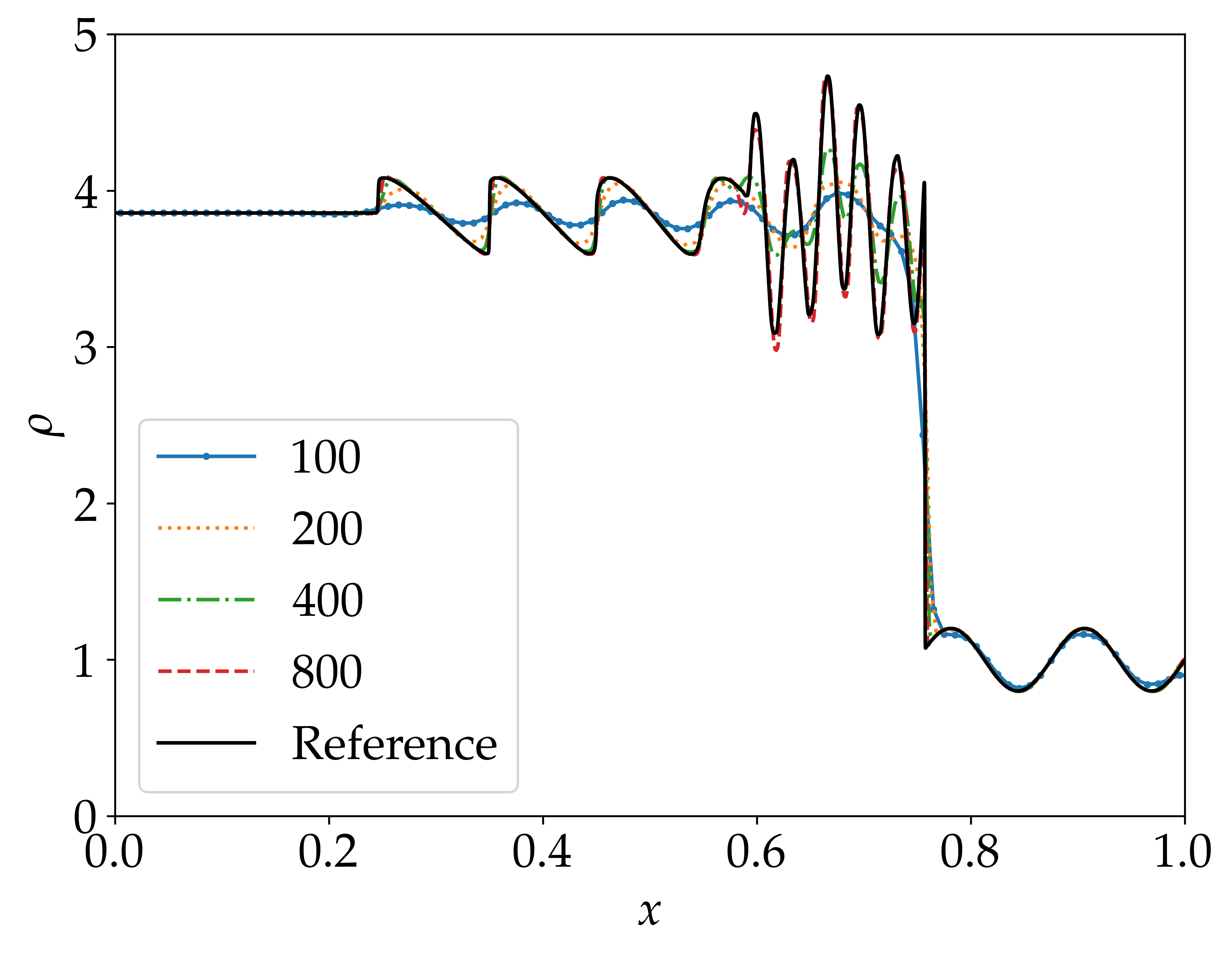}}
    \caption{Numerical results of density $\rho$ of the 1D Euler equations with Shu-Osher shock tube initial condition under different mesh sizes generated from (a) RLS-D and (b) RLS-A.}
  \label{fig:euler3.2}
\end{figure}

In the Shu-Osher problem test, as shown in Fig.~\ref{fig:euler3.2}a, RLS-D is not able to predict the high-frequency waves caused by shock-entropy wave interaction even with 800 cells. Instead, RLS-A captures the high-frequency waves more sharply although there exist certain density value overshoots in the high-frequency wave region when 800 cells are used (see Fig.~\ref{fig:euler3.2}b).

\subsection{Generalization across different problem dimensions}
\label{2d}
In this section, the same schemes trained from the 1D Burger's equation are used to solve the 2D Euler equations with tensor-product-based solution and flux constructions.
In the 2D form, Euler equations consist of four equations that govern the conservation of mass, momentum and energy. This reads
\begin{equation}
\frac{\partial U}{\partial t} + \frac{\partial F}{\partial x} + \frac{\partial G}{\partial y}= 0,
\label{eq:Euler_2D}
\end{equation}
with
\begin{eqnarray*}
U  = 
\left (
\begin{array}{c}
      \rho   \\
      \rho u \\
      \rho v \\
      E
\end{array} \right ), \quad 
F =
\left (
\begin{array}{c}
      \rho u  \\
      \rho u^2 + p  \\
      \rho u v \\
      u (E+p)
\end{array} \right ), \quad \text{and} \quad
G =
\left (
\begin{array}{c}
      \rho v  \\
      \rho u v \\
      \rho v^2 + p  \\
      v (E+p)
\end{array} \right ).
\end{eqnarray*}

Herein, $\rho$ is the density, $u$ and $v$ are the velocity components in the $x$ and $y$ directions, respectively, $p$ is the pressure, and $E$ is the total energy. The system is closed with the perfect gas law.

Two 2D Riemann problems are simulated in this section. The computational domain $[0,1] \times [0,1]$ is divided into four quadrants centered at $(0.5,0.5)$. The initial conditions of $p$, $\rho$, $u$, and $v$ in different quadrants for the two problems are given in Table~\ref{tab:configuration}.

\begin{table}[!htb]
\centering
\begin{tabular}{|c|c|c|}
\hline
\multirow{2}{*}{IC1} & $p_2 = 1$,  $\rho_2 = 1$,        $u_2 = 0.7276$,   $v_2 = 0$ & $p_1 = 0.4$,          $\rho_1 = 0.5313$,  $u_1 = 0$,  $v_1 = 0$  \\       \cline{2-3} 
    & $p_3 = 1$,  $\rho_3 = 0.8$, $u_3 = 0$, $v_3 = 0$ & $p_4 = 1$,  $\rho_4 = 1$,  $u_4 = 0$,   $v_4 = 0.7276$  \\ 
\hline
\hline
\multirow{2}{*}{IC2} &  $p_2 = 1$,  $\rho_2 = 2$,  $u_2 = 0.75$,   $v_2 = 0.5$  & $p_1 = 1$,  $\rho_1 = 1$,  $u_1 = 0.75$,  $v_1 = -0.5$ \\ \cline{2-3} 
                     & $p_3 = 1$,  $\rho_3 = 1$,  $u_3 = -0.75$,  $v_3 = 0.5$ & $p_4 = 1$,  $\rho_4 = 3$,  $u_4 = -0.75$,   $v_4 = -0.5$  \\ \hline
\end{tabular}
\caption{Initial conditions in different quadrants for Case 1 and Case 2}
\label{tab:configuration}
\end{table}

\begin{figure}[!htbp]
  \centering
  \includegraphics[width=1.0\textwidth]{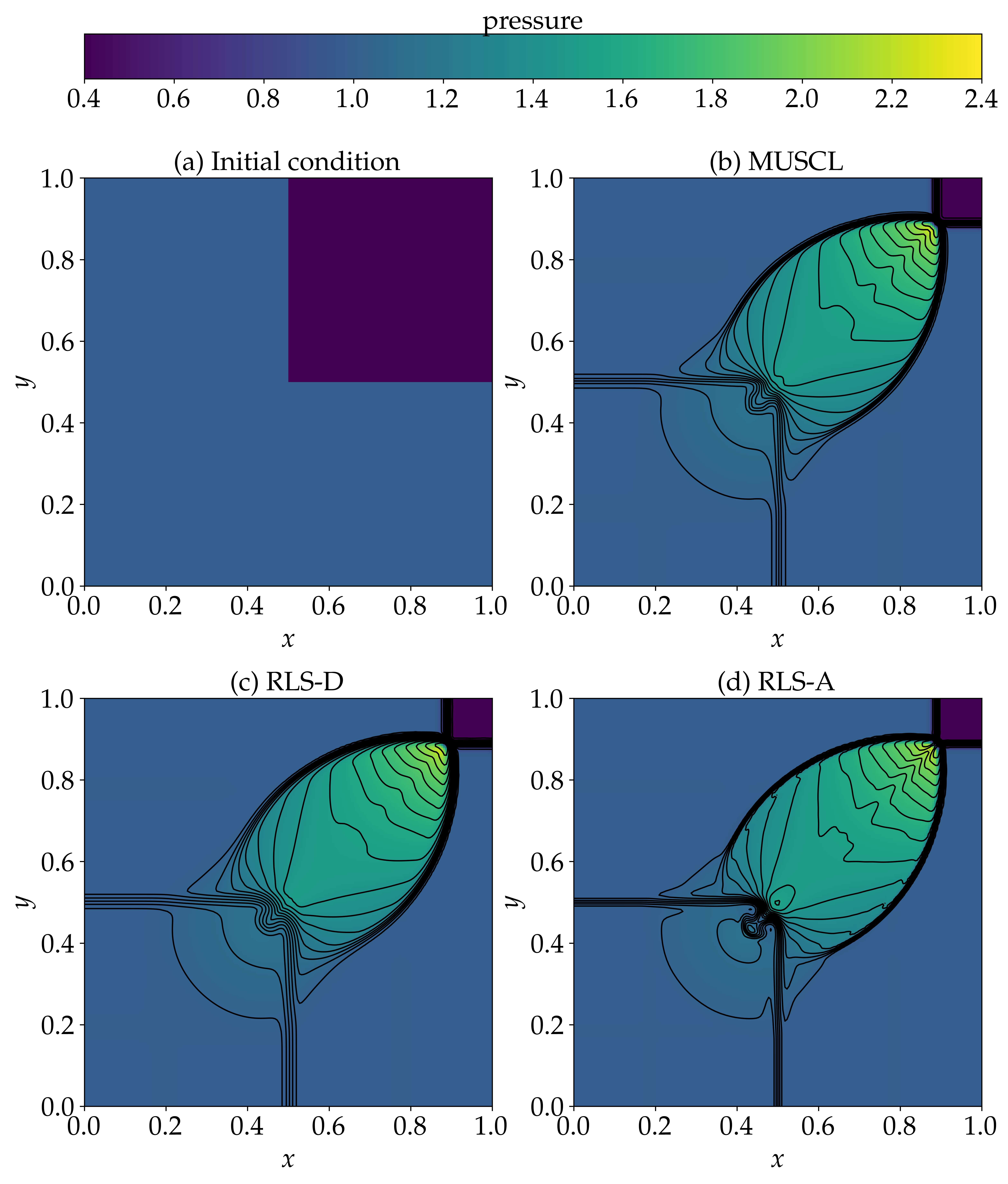} 
    \caption{Numerical results of 2D Euler equations with Case 1 (a) initial condition generated from (b) the third-order MUSCL scheme with van Albada limiter and (c) RLS-D, (d) RLS-A. Pressure is displayed by color and density by 30 contours (0.54 to 1.7 step 0.04)~\cite{liska2003comparison}.}
  \label{fig:2d-euler1}
\end{figure}

\begin{figure}[!htbp]
  \centering
  \includegraphics[width=1.0\textwidth]{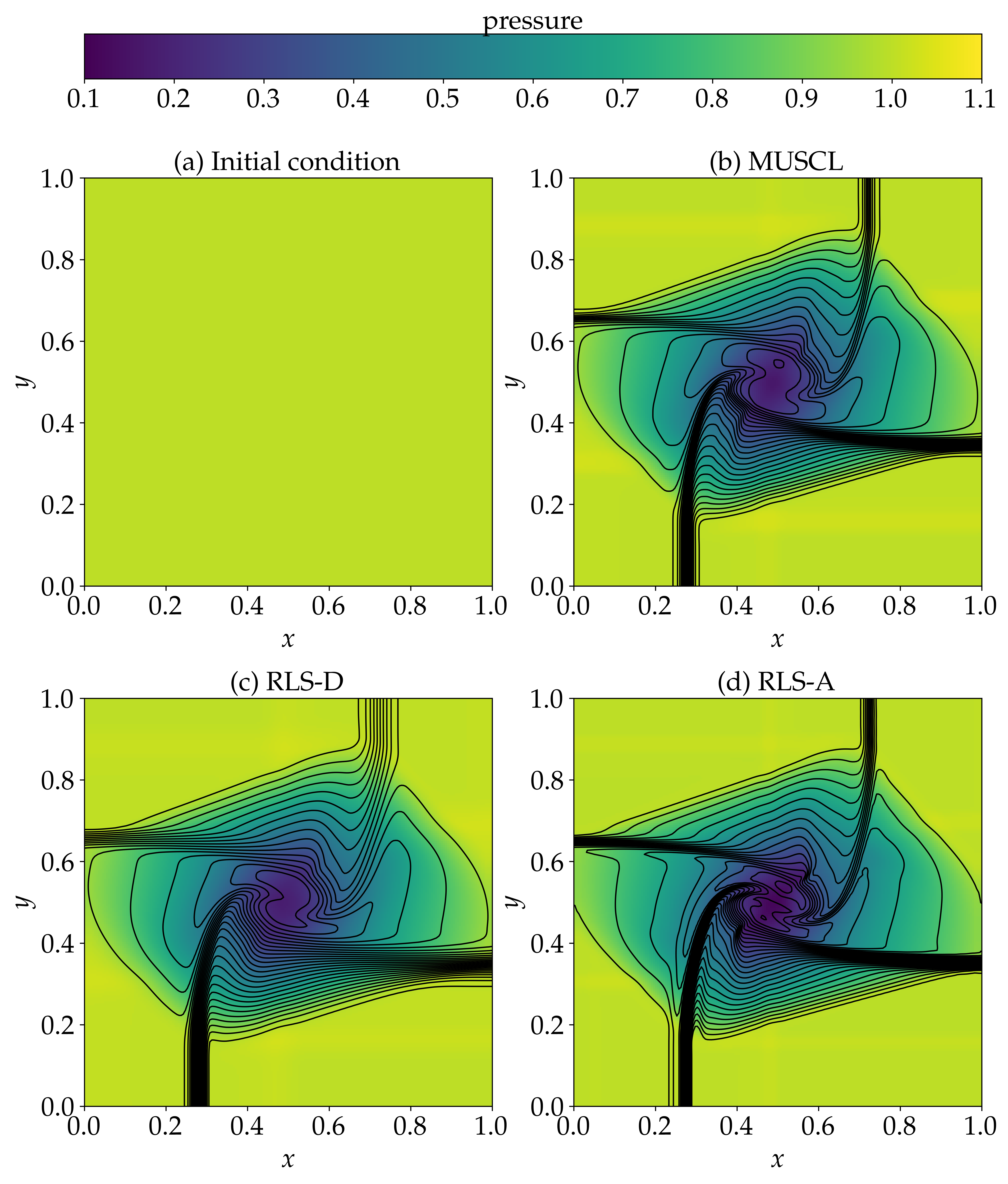} 
    \caption{Numerical results of 2D Euler equations with Case 2 (a) initial condition generated from (b) the third-order MUSCL scheme with van Albada limiter and (c) RLS-D, (d) RLS-A. Pressure is displayed by color and density by 29 contours (0.25 to 3.05 step 0.1).}
  \label{fig:2d-euler2}
\end{figure}

The grid consists of $200 \times 200$ cells in the $x$ and $y$ directions, respectively. The CFL number is $0.1$, and the simulation end time is $T_{\textrm{end}}  = 0.25$ for Case 1 and $T_{\textrm{end}}  = 0.3$ for Case 2. In the tensor-product-based approach, directional spatial splitting is applied in each spatial dimension and the method of lines is used to march the resulting system of ordinary differential equations in time. Thus, the 1D numerical schemes trained from Burger's equation can be directly used in the 2D implementation. As in previous tests, the second-order Runge–Kutta method is used to march the system in time. 

We observe from both Figs.~\ref{fig:2d-euler1} and \ref{fig:2d-euler2} that simulation results from the trained numerical schemes RLS-A and RLS-D agree reasonably well with those calculated from the third-order MUSCL scheme with van Albada limiter. The results generated from RLS-A show sharper resolutions of flow discontinuities and vortex structures. Instead, RLS-D shows large numerical dissipation features. As shown in Fig.~\ref{fig:2d-euler2}c, RLS-D can significantly smear the contact discontinuity, similar to that observed in Fig.~\ref{fig:euler3.1}a from 1D simulations.

\subsection{Tests of extreme flow conditions}
\begin{figure}[!htbp]
  \centering
  {\includegraphics[width=0.7\textwidth]{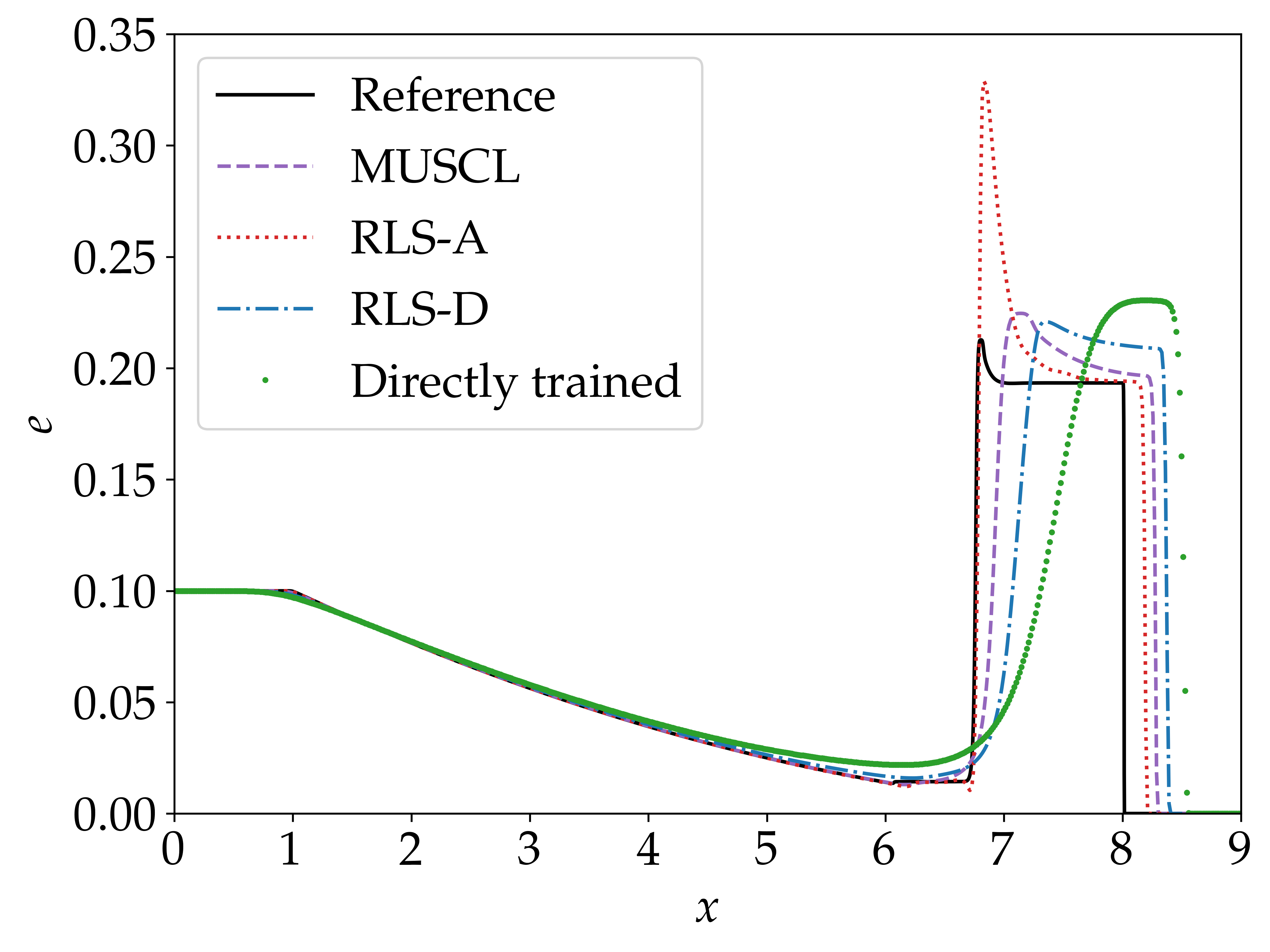}}
    \caption{Comparison of numerical simulation results of internal energy $e$ of the 1D Euler equations with Leblanc shock tube initial condition generated from RLS-A, RLS-D, scheme directly trained from the Leblanc environment, the third-order MUSCL scheme with van Albada limiter and reference solution.}
  \label{fig:leblanc}
\end{figure}

We note that the MARL-based nonlinear numerical schemes learned from Burger's equation cannot recognize positivity constraints on physical variables, such as density $\rho$, pressure $p$ and internal energy $e$, from Euler equations. Therefore, it is expected that the so-far well-behaved nonlinear numerical schemes RLS-A and RLS-D may encounter difficulty when applied to shock capturing under extreme flow conditions.
Here we tested the new MARL-based nonlinear numerical schemes with the 1D Leblanc shock tube problem with extreme internal energy jump. The initial conditions for density $\rho$, velocity $u$, and specific internal energy $e$ are given as follows:
\begin{equation}
    (\rho,\, u,\, e) = \Bigl\{
    \begin{array}{lcl}(1.0,\,0,\,0.1) & x<3\\
    (0.001,\,0.0,\,10^{-7}) & x\ge3
    \end{array}.
\end{equation}
The constant specific heat ratio $\zeta$ is $5/3$, the domain of interest is $[0, 9]$, the number of cells is $N_x=600$, the simulation end time is $T_{\textrm{end}} = 6$ and the CFL number is $0.1$.  The reference solution is calculated using the same MUSCL scheme on a finer grid, where the number of cells is $N_x=4800$.
We found that when RLS-A and RLS-D trained from Burger's equation simulations were used to solve the Leblanc shock tube problem, simulations blew up due to the creation of negative internal energy. To solve this issue, two approaches are used. In the first approach, a positivity limiting procedure inspired by the work of Zhang and Shu~\cite{zhang2010positivity} is used to regulate the outputs of RLS-A and RLS-D. In the second, the MARL-based scheme was directly trained in the Leblanc shock tube environment. The results are presented in Fig.~\ref{fig:leblanc}. We observed that both approaches can solve the positivity issue. RLS-A with the positivity limiter has the smallest numerical dissipation, and the directly trained MARL-based scheme has the largest numerical dissipation. The inferior performance of the directly trained MARL-based scheme is caused by the excessive numerical dissipation preferred by the training process. Since any overshoot in the internal energy or density during the early simulation stages will make the solver blow up due to the creation of negative internal energy or density, the interaction between the agents and environment steers the action control policy towards preferring stability over accuracy. Thus, satisfying the TVD property dominates the reward~\eqref{eq:reward}, and large numerical dissipation is then preferred by the reinforcement learning model. Simply encoding positivity properties into the reward does not work for this challenging case as the dominating solution to ensure positivity is to strictly enforce TVD. That is the reason why in the first approach the positivity limiting procedure is used to regulate the outputs of MARL-based schemes as an extra post-processing operation.

\section{Conclusions}
\label{conclusion}
This work presents a universal, first-principle-like MARL-based framework to design nonlinear numerical schemes for general hyperbolic conservation laws. Different from existing reinforcement learning based research, no reference data or empirical elements are incorporated in the reward design. Instead, the reward function is formulated based on a first-principle-like approach using the fundamental CFD theories. Specifically, numerical stability is achieved by controlling the total variation of numerical solutions, and numerical accuracy is promoted by using $k$-exact polynomial reconstruction of solution variables. The agents are able to strike a balance between accuracy and numerical dissipation of the learned numerical schemes through iterative  interaction with the governing equations (i.e., environment in reinforcement learning) and exploitation of many interaction samples. The 1D inviscid Burger's equation is used to train the MARL-based nonlinear numerical schemes. Then the generalizability of the learned schemes is tested with both 1D and 2D Euler equation simulations admitting shock waves under different flow conditions and grid resolutions.
The MARL-based schemes are able to obtain comparable results to those using the third-order MUSCL scheme equipped with the van Albada limiter, and the learned scheme with low numerical dissipation (i.e., RLS-A) outperforms the MUSCL scheme. These promising numerical simulation results demonstrate a new paradigm for designing nonlinear numerical schemes by using reinforcement learning with first-principle-like rewards.

From tests of extreme flow conditions with the 1D Leblanc shock tube problem, we find that when interaction samples used in reinforcement learning favor certain reward criteria, e.g., the TVD property for scheme stability due to harsh flow conditions, the learned numerical scheme can be biased, e.g., the scheme is very dissipative when scheme stability is overly emphasized. There is no easy solution of such issues as interaction samples favoring other reward criteria, e.g., $k$-exact reconstruction for scheme accuracy in the Leblanc problem, are rare. To remedy this situation, post-processing the learned numerical scheme with domain knowledge provides an alternative solution. Since it is known that negative density and internal energy can be easily generated when simulating the Leblanc problem, accuracy-preserving positivity limiting procedures, such as those developed by Zhang and Shu~\cite{zhang2010positivity}, can be used to regulate the solution behavior. 
Note that simply encoding positivity requirements into the reward does not create desired agents with low numerical dissipation as enforcement of the TVD property can avoid the creation of negative density and internal energy. More sophisticated interaction sample selection mechanism may need to be developed for reinforcement learning of flows with extreme conditions.

\section*{Acknowledgments}
HW and HX are funded by Deutsche Forschungsgemeinschaft (DFG, German Research Foundation) under Germany's Excellence Strategy - EXC 2075 – 390740016. We acknowledge the support by the Stuttgart Center for Simulation Science (SimTech).

\appendix
\section{Agent Training Methods in Reinforcement Learning}
\label{sec:TD3}
The goal of reinforcement learning is to find an optimal policy that guides an agent in choosing actions to maximize the expected return. 
There are two main approaches to train agents: policy-based and value-based. Policy-based methods represent the policy explicitly and the training is much more stable and reliable because these methods directly optimize for the policy. On the other hand, value-based methods search for the optimal action-value function and indirectly optimize for agent performance which can lead to less stable training. Nevertheless, value-based methods are more sample-efficient compared to the policy-based methods.

To take full advantage of the aforementioned two agent training approaches, a new family of methods combining policy-based and value-based methods, known as actor-critic methods, were developed~\cite{konda1999actor}. In this framework, an actor controls how the agent behaves and a critic evaluates the Q value of a certain action.
Deep Deterministic Policy Gradient (DDPG) is one of this type of methods that concurrently learns a Q-learning function and a policy~\cite{lillicrap2015continuous}.
A Q-learning function describes the expected return after taking an action $a$ in state $s$ and therefore following policy $\pi$:
\begin{equation}
Q_{\theta}(s,a) = r + \gamma \mathbb{E}_{s',a'}[Q_{\theta}(s',a')], a'\sim \pi(s').
\end{equation}
The algorithm optimizes the Q-learning function by minimizing the loss:
\begin{equation}
L(\psi) = \mathbb{E} [(Q_{\theta}(s,a) - y(r,s'))^2],
\end{equation}
where
\begin{equation}
y(r,s') = r + \gamma Q(s',a').
\end{equation}
DDPG introduces two approaches to improve the performance: the use of a replay buffer, and target networks. A replay buffer contains a certain amount of previous transition tuples and is used in almost every deep reinforcement learning algorithm to make the training stable. When the algorithm minimizes the loss, it is trying to make the action-value function be more like the target $y_t$. However, the target depends on the same parameters $\psi$ the algorithm is going to train which again makes the training unstable. The solution is to create a copy of the actor and critic neural networks and mark them as the target networks. These target networks are updated more slowly than the main networks which greatly improves the stability of learning, \(\theta' \leftarrow \tau \theta + (1-\tau)\theta'\). 

Twin delayed deep deterministic policy gradient (TD3) further improves the performance of DDPG by introducing three new numerical treatments: clipped double Q-learning functions, delayed policy updates, and target policy smoothing~\cite{fujimoto2018addressing}.
Clipped double Q-learning is used to ease the problem of the overestimation of Q-values. Two critic neural networks $Q_{\theta_1}, Q_{\theta_2}$ are defined and the algorithm updates the target by always taking the minimum between the two estimates:
\begin{equation}
y_1 = r + \gamma \mbox{min}_{k=1,2} Q_{\theta'_k}(s',\tilde{a}).
\end{equation}

To mitigate the problem of policy updates with a high variance value estimate, the policy network and the target network are updated at a lower frequency than the value network. This will limit the likelihood of repeating updates with respect to an unchanged critic. We use the same update frequency as the original suggestion, i.e., one policy update every two critic updates. Moreover, to ensure the error remains small, the target networks are updated slowly as in the DDPG.

The third numerical treatment, target policy smoothing regularization, is introduced to avoid the exploitation of the peak in the value estimate by adding a small amount of random noise to the target action: 
\begin{equation}
\tilde{a} \leftarrow \mbox{clip}  (\pi_{\psi'}(s') + \mbox{clip}(\mathcal{N}(0,\sigma),-c,c),a_{\textrm{low}},a_{\textrm{high}}).
\end{equation}

With the aforementioned three numerical treatments included in the algorithm, TD3 greatly improves both the learning speed and performance of DDPG in the continuous control setting and makes TD3 a suitable reinforcement learning algorithm for our problem.

\section{Parameter Selection in Reward Design}
\label{sec:parameter}
The idea behind constructing the blending function $\alpha$ in Eqn.~\eqref{eq:reward} is that the algorithm needs to reach a good balance between numerical dissipation and accuracy. Specifically, $\alpha \, r\textsubscript{D}$ should have a similar order of magnitude to that of $r\textsubscript{A}$. From numerical experiments, we indeed observed that $\alpha \, r\textsubscript{D}$ and $r\textsubscript{A}$ exhibit similar orders of magnitude for each cell when a good balance is reached. Since under the same flow conditions the CFL number is positively correlated to the change of total variation, we approximate $\alpha$ as $\alpha_0/\text{CFL}$ to automatically control the weight of $r\textsubscript{D}$ for varying CFL numbers in this study. 

Note that a constant $\alpha_0$ is introduced when constructing the blending function $\alpha$. At a first glance, it seems that $\alpha_0$ can be a random number as the ratio between $r\textsubscript{A}$ and $r\textsubscript{D}$ depends on the problem to be solved. However, a preliminary analysis indicates that the order of magnitude of $\alpha_0$ can be estimated when the flow problem is appropriately non-dimensionalized. 
The estimation procedure can go as follows. Based on the definition of $r\textsubscript{A}$, it maximum magnitude is $7/3$, around $\mathcal{O} (1)$. 
Therefore, the order of magnitude of $\alpha_0 \, r\textsubscript{D}/\text{CFL}$ should be targeted around $\mathcal{O} (1)$ for all problems as the order of magnitude of $r\textsubscript{A}$ always holds. Since $r\textsubscript{A}$, $r\textsubscript{D}$, and CFL all have the dimension of one, the dimension of $\alpha_0$ also needs to be one. At this point, numerical experiments can be carried out with non-dimensionalized problems to determine the order of magnitude of $r\textsubscript{D}$. 
Before moving forward, we mention that the order of magnitude of $r\textsubscript{D}$ can actually be estimated from numerical simulations with the MUSCL scheme. Therein, we found that $r\textsubscript{D}$ is around $\mathcal{O} (10^{-2})$, and thus, $\alpha_0/\text{CFL}$ should be around $\mathcal{O} (10^{2})$ to make the order of magnitude of $\alpha \, r\textsubscript{D}$ be around $\mathcal{O} (1)$, matching that of $ r\textsubscript{A}$. Since the CFL number varies between $\mathcal{O} (10^{-1})$ and $ \mathcal{O} (1)$, the value of $\alpha_0$ can be between $\mathcal{O} (10)$ and $ \mathcal{O} (10^2)$.

Now we present the observations from our numerical experiments. For all simulations, the CFL number is fixed at $0.2$. 
We varied $\alpha_0$ in a wide range from $\mathcal{O} (10^{-1})$ to $\mathcal{O} (10^3)$, and found that the total variation change between two successive time steps for appropriately non-dimensionalized problems shows sloppy features. This sloppiness actually ensures the effectiveness of numerical experiments used to estimate the order of magnitude of $r\textsubscript{D}$ and the value of $\alpha_0$.
A key observation is that the reinforcement learning algorithm can always learn a certain policy to maximize the reward. However, the balance between $r\textsubscript{D}$ and $r\textsubscript{A}$ can be undesirable due to different choices of $\alpha_0$. If $\alpha_0$ is set as a large number (e.g., $10^{3}$), the penalty for violating the TVD constraint can be very severe (i.e., $\alpha \, r\textsubscript{D}$ is a large negative number). The algorithm then simply chose actions that satisfy the TVD constraints leading to highly dissipative numerical schemes. 
If $\alpha_0$ is set as a small number (e.g., $0.1$), we observed that the final policy excessively favored accuracy and led to spurious oscillations in simulation results because $r_A$ dominated in the reward. From extensive tests, $\alpha_0$ between 1 and 100 creates good balance between numerical dissipation and accuracy. Therefore, $\alpha_0 = 50$ is used in nonlinear numerical scheme training, and the machine-learned schemes can be generalized to different physics, grid resolutions, and spatial dimensions, as demonstrated in Section~\ref{sec:result}.



\begin{thebibliography}{10}
\expandafter\ifx\csname url\endcsname\relax
  \def\url#1{\texttt{#1}}\fi
\expandafter\ifx\csname urlprefix\endcsname\relax\def\urlprefix{URL }\fi
\expandafter\ifx\csname href\endcsname\relax
  \def\href#1#2{#2} \def\path#1{#1}\fi

\bibitem{meister2012hyperbolic}
A.~Meister, J.~Struckmeier, Hyperbolic Partial Differential Equations: Theory,
  Numerics and Applications, Springer Science \& Business Media, 2012.

\bibitem{godunov1959finite}
S.~K. Godunov, I.~Bohachevsky, Finite difference method for numerical
  computation of discontinuous solutions of the equations of fluid dynamics,
  Matemati{\v{c}}eskij sbornik 47~(3) (1959) 271--306.

\bibitem{van1979towards}
B.~Van~Leer, Towards the ultimate conservative difference scheme. {V}. {A}
  second-order sequel to {G}odunov's method, Journal of Computational Physics
  32~(1) (1979) 101--136.

\bibitem{harten1997uniformly}
A.~Harten, B.~Engquist, S.~Osher, S.~R. Chakravarthy, Uniformly high order
  accurate essentially non-oscillatory schemes, III, Springer, 1997.

\bibitem{jiang1996efficient}
G.-S. Jiang, C.-W. Shu, Efficient implementation of weighted {ENO} schemes,
  Journal of Computational Physics 126~(1) (1996) 202--228.

\bibitem{cockburn1998essentially}
B.~Cockburn, C.-W. Shu, C.~Johnson, E.~Tadmor, C.-W. Shu, Essentially
  non-oscillatory and weighted essentially non-oscillatory schemes for
  hyperbolic conservation laws, Springer, 1998.

\bibitem{Cockburn_Shu_1989MC}
B.~Cockburn, C.-W. Shu, {TVB} {R}unge-{K}utta local projection discontinuous
  {G}alerkin finite element method for conservation laws. {II}. {G}eneral
  framework, Mathematics of Computation 52~(186) (1989) 411--435.

\bibitem{kim2005accurate}
K.~H. Kim, C.~Kim, Accurate, efficient and monotonic numerical methods for
  multi-dimensional compressible flows: Part {II}: Multi-dimensional limiting
  process, Journal of Computational Physics 208~(2) (2005) 570--615.

\bibitem{you2018high}
H.~You, C.~Kim, High-order multi-dimensional limiting strategy with subcell
  resolution {I}. {T}wo-dimensional mixed meshes, Journal of Computational
  Physics 375 (2018) 1005--1032.

\bibitem{corrigan2019moving}
A.~Corrigan, A.~D. Kercher, D.~A. Kessler, A moving discontinuous {G}alerkin
  finite element method for flows with interfaces, International Journal for
  Numerical Methods in Fluids 89~(9) (2019) 362--406.

\bibitem{luo2021moving}
H.~Luo, G.~Absillis, R.~Nourgaliev, A moving discontinuous {G}alerkin finite
  element method with interface condition enforcement for compressible flows,
  Journal of Computational Physics 445 (2021) 110618.

\bibitem{persson2006sub}
P.-O. Persson, J.~Peraire, {Sub-cell shock capturing for discontinuous Galerkin
  methods}, in: 44th AIAA Aerospace Sciences Meeting and Exhibit, 2006, p. 112.

\bibitem{yu2015localized}
M.~Yu, F.~X. Giraldo, M.~Peng, Z.-J. Wang, Localized artificial viscosity
  stabilization of discontinuous {G}alerkin methods for nonhydrostatic
  mesoscale atmospheric modeling, Monthly Weather Review 143~(12) (2015)
  4823--4845.

\bibitem{haga2019robust}
T.~Haga, S.~Kawai, On a robust and accurate localized artificial diffusivity
  scheme for the high-order flux-reconstruction method, Journal of
  Computational Physics 376 (2019) 534--563.

\bibitem{ray2018artificial}
D.~Ray, J.~S. Hesthaven, An artificial neural network as a troubled-cell
  indicator, Journal of Computational Physics 367 (2018) 166--191.

\bibitem{ray2019detecting}
D.~Ray, J.~S. Hesthaven, Detecting troubled-cells on two-dimensional
  unstructured grids using a neural network, Journal of Computational Physics
  397 (2019) 108845.

\bibitem{beck2020neural}
A.~D. Beck, J.~Zeifang, A.~Schwarz, D.~G. Flad, A neural network based shock
  detection and localization approach for discontinuous {G}alerkin methods,
  Journal of Computational Physics 423 (2020) 109824.

\bibitem{nguyen2022machine}
N.~Nguyen-Fotiadis, M.~McKerns, A.~Sornborger, Machine learning changes the
  rules for flux limiters, Physics of Fluids 34~(8) (2022).

\bibitem{bar2019learning}
Y.~Bar-Sinai, S.~Hoyer, J.~Hickey, M.~P. Brenner, Learning data-driven
  discretizations for partial differential equations, Proceedings of the
  National Academy of Sciences 116~(31) (2019) 15344--15349.

\bibitem{silver2017mastering}
D.~Silver, J.~Schrittwieser, K.~Simonyan, I.~Antonoglou, A.~Huang, A.~Guez,
  T.~Hubert, L.~Baker, M.~Lai, A.~Bolton, et~al., Mastering the game of go
  without human knowledge, Nature 550~(7676) (2017) 354--359.

\bibitem{mnih2013playing}
V.~Mnih, K.~Kavukcuoglu, D.~Silver, A.~Graves, I.~Antonoglou, D.~Wierstra,
  M.~Riedmiller, Playing {A}tari with deep reinforcement learning, arXiv
  preprint arXiv:1312.5602 (2013).

\bibitem{zhao2020sim}
W.~Zhao, J.~P. Queralta, T.~Westerlund, Sim-to-real transfer in deep
  reinforcement learning for robotics: a survey, in: 2020 IEEE Symposium Series
  on Computational Intelligence (SSCI), IEEE, 2020, pp. 737--744.

\bibitem{hu2023remedi}
C.~Hu, K.~V. Saboo, A.~H. Ali, B.~D. Juran, K.~N. Lazaridis, R.~K. Iyer,
  {REMEDI}: {RE}inforcement learning-driven adaptive {ME}tabolism modeling of
  primary sclerosing cholangitis {DI}sease progression, arXiv preprint
  arXiv:2310.01426 (2023).

\bibitem{yu2021reinforcement}
C.~Yu, J.~Liu, S.~Nemati, G.~Yin, Reinforcement learning in healthcare: A
  survey, ACM Computing Surveys (CSUR) 55~(1) (2021) 1--36.

\bibitem{chen2023deep}
W.~Chen, Q.~Wang, L.~Yan, G.~Hu, B.~R. Noack, Deep reinforcement learning-based
  active flow control of vortex-induced vibration of a square cylinder, Physics
  of Fluids 35~(5) (2023).

\bibitem{novati2021automating}
G.~Novati, H.~L. de~Laroussilhe, P.~Koumoutsakos, Automating turbulence
  modelling by multi-agent reinforcement learning, Nature Machine Intelligence
  3~(1) (2021) 87--96.

\bibitem{bae2022scientific}
H.~J. Bae, P.~Koumoutsakos, Scientific multi-agent reinforcement learning for
  wall-models of turbulent flows, Nature Communications 13~(1) (2022) 1443.

\bibitem{sutton2018reinforcement}
R.~S. Sutton, A.~G. Barto, Reinforcement learning: An introduction, MIT press,
  2018.

\bibitem{wang2019learning}
Y.~Wang, Z.~Shen, Z.~Long, B.~Dong, Learning to discretize: solving 1{D} scalar
  conservation laws via deep reinforcement learning, arXiv preprint
  arXiv:1905.11079 (2019).

\bibitem{way2022backpropagation}
E.~Way, D.~S. Kapilavai, Y.~Fu, L.~Yu, Backpropagation through time and space:
  learning numerical methods with multi-agent reinforcement learning, arXiv
  preprint arXiv:2203.08937 (2022).

\bibitem{fu2022multi}
Y.~Fu, D.~S. Kapilavai, E.~Way, Multi-agent learning of numerical methods for
  hyperbolic {PDE}s with factored {D}ec-{MDP}, in: International Conference on
  Practical Applications of Agents and Multi-Agent Systems, Springer, 2022, pp.
  179--190.

\bibitem{fu2016family}
L.~Fu, X.~Y. Hu, N.~A. Adams, A family of high-order targeted {ENO} schemes for
  compressible-fluid simulations, Journal of Computational Physics 305 (2016)
  333--359.

\bibitem{feng2023deep}
Y.~Feng, F.~S. Schranner, J.~Winter, N.~A. Adams, A deep reinforcement learning
  framework for dynamic optimization of numerical schemes for compressible flow
  simulations, Journal of Computational Physics 493 (2023) 112436.

\bibitem{schwarz2023reinforcement}
A.~Schwarz, J.~Keim, S.~Chiocchetti, A.~Beck, A reinforcement learning based
  slope limiter for second-order finite volume schemes, PAMM 23~(1) (2023)
  e202200207.

\bibitem{keim2023reinforcement}
J.~Keim, A.~Schwarz, S.~Chiocchetti, C.~Rohde, A.~Beck, A reinforcement
  learning based slope limiter for two-dimensional finite volume schemes, in:
  International Conference on Finite Volumes for Complex Applications,
  Springer, 2023, pp. 209--217.

\bibitem{fujimoto2018addressing}
S.~Fujimoto, H.~Hoof, D.~Meger, Addressing function approximation error in
  actor-critic methods, in: International Conference on Machine Learning, PMLR,
  2018, pp. 1587--1596.

\bibitem{paszke2019pytorch}
A.~Paszke, S.~Gross, F.~Massa, A.~Lerer, J.~Bradbury, G.~Chanan, T.~Killeen,
  Z.~Lin, N.~Gimelshein, L.~Antiga, et~al., Pytorch: An imperative style,
  high-performance deep learning library, Advances in Neural Information
  Processing Systems 32 (2019).

\bibitem{brockman2016openai}
G.~Brockman, V.~Cheung, L.~Pettersson, J.~Schneider, J.~Schulman, J.~Tang,
  W.~Zaremba, Open{AI} {G}ym, arXiv preprint arXiv:1606.01540 (2016).

\bibitem{agarap2018deep}
A.~F. Agarap, Deep learning using rectified linear units ({R}e{LU}), arXiv
  preprint arXiv:1803.08375 (2018).

\bibitem{kingma2014adam}
D.~P. Kingma, J.~Ba, Adam: A method for stochastic optimization, arXiv preprint
  arXiv:1412.6980 (2014).

\bibitem{liska2003comparison}
R.~Liska, B.~Wendroff, Comparison of several difference schemes on 1{D} and
  2{D} test problems for the {E}uler equations, SIAM Journal on Scientific
  Computing 25~(3) (2003) 995--1017.

\bibitem{zhang2010positivity}
X.~Zhang, C.-W. Shu, On positivity-preserving high order discontinuous
  {G}alerkin schemes for compressible {E}uler equations on rectangular meshes,
  Journal of Computational Physics 229~(23) (2010) 8918--8934.

\bibitem{konda1999actor}
V.~Konda, J.~Tsitsiklis, Actor-critic algorithms, Advances in neural
  information processing systems 12 (1999).

\bibitem{lillicrap2015continuous}
T.~P. Lillicrap, J.~J. Hunt, A.~Pritzel, N.~Heess, T.~Erez, Y.~Tassa,
  D.~Silver, D.~Wierstra, Continuous control with deep reinforcement learning,
  arXiv preprint arXiv:1509.02971 (2015).

\end{thebibliography}
\end{document}